\documentclass[lettersize,journal]{IEEEtran}
\usepackage{amsmath,amsfonts}
\usepackage{algorithmic}
\usepackage{booktabs}
\usepackage{algorithm}
\usepackage{array}

\usepackage{textcomp}
\usepackage{stfloats}
\usepackage{url}
\usepackage{verbatim}
\usepackage{graphicx}
\usepackage{cite}
\usepackage{graphicx}
\usepackage{subcaption}

\newtheorem{problem}{Problem}
\newtheorem{remark}{Remark}

\def\BibTeX{{\rm B\kern-.05em{\sc i\kern-.025em b}\kern-.08em
    T\kern-.1667em\lower.7ex\hbox{E}\kern-.125emX}}
\begin{document}
\title{Differentiable Hybrid-Action Neural Feedback Control for District Heating Networks}
\author{Nicolas Kirsch, Corrado Sgadari, Alessio La Bella, Giancarlo Ferrari-Trecate
\thanks{This research has been supported by the Swiss National Sci-
ence Foundation under the NCCR Automation (grant agreement
51NF40 180545) and the NECON project (grant number 200021-
219431).}}

\maketitle

\begin{abstract}
Many cyber-physical systems require control policies that combine continuous setpoints with discrete operational decisions, such as equipment switching, mode selection, or resource scheduling. Discrete actions are not differentiable, which obstructs gradient-based policy training, while conventional mixed-integer formulations remain costly to solve online. This paper proposes a hybrid-action neural controller (HANC) in which a continuous branch, a categorical branch and a differentiable assembly layer jointly generate  commands that satisfy complex actuator constraints by construction. Categorical decisions are handled using a straight-through Gumbel estimator, enabling the policy to be trained by backpropagation through time over full closed-loop rollouts.

The proposed framework is deployed on a district heating network (DHN) featuring multiple heat generation units and stratified thermal energy storage. Its performance is evaluated on a simulation of a real DHN located at RSE SpA in Italy. The resulting policy jointly learns switching decisions and continuous operating setpoints. Under dynamic electricity pricing, the learned controller reduces operating cost by $\mathbf{30\%}$ compared to a rule-based industrial baseline. 
We also show that, compared with a deterministic straight-through relaxation, injecting noise during training achieves similar cost while reducing hard switching by an order of magnitude, and attribute this difference to the wider decision margins of the resulting policy.

\end{abstract}


\section{Introduction}
\label{sec:introduction}

Modern control systems increasingly rely on the joint use of categorical and continuous control actions. Hybrid decision-making problems arise whenever controllers must simultaneously determine discrete operating modes and continuous control inputs. Coordinating these decisions over time is challenging because they are tightly coupled through the system dynamics, making the synthesis of high-performance control policies difficult.

This challenge is particularly pronounced in modern energy systems, including power systems and energy communities, and is especially relevant to district heating networks (DHNs). Their operation requires balancing multiple objectives, such as economic efficiency, reliability, and environmental performance. At the same time, these systems comprise heterogeneous assets with distinct operating characteristics, requiring controllers to determine both which units should be active and their corresponding operating setpoints~\cite{MANCARELLA20141,blanco2019novel}. Moreover, the operational constraints of many devices introduce discrete operating modes and on/off decisions alongside continuous actuation variables~\cite{hering2022operational}. The nonlinear network dynamics and multiple objectives, combined with the mixed-integer nature of the control problem, make optimal control of DHNs particularly challenging. This combination calls for control methods that can jointly handle network dynamics and mixed-integer decisions while remaining computationally efficient for practical deployment.

\subsection{Related work}

Industrial practice in DHN operation remains dominated by simple, robust heuristics \cite{app11010455}. Supply temperatures are commonly determined through fixed setpoints or outdoor-temperature compensation curves \cite{SLEPTSOV2021107598}, while pumps and generation units are operated according to schedules or rule-based logic encoding operator experience \cite{nigro2025}. Such strategies remain widespread due to their simplicity, reliability, and limited requirements in terms of system modelling, forecasting, and computational resources. By relying on predefined operating rules, they effectively circumvent the computational complexity associated with the hybrid nature of DHN operation. However, this simplicity comes at the cost of reduced adaptability and optimality. Their reactive and fixed structure limits their ability to exploit future information, such as electricity price variations or anticipated demand changes, leading to missed arbitrage opportunities and increased operating costs.

Beyond rule-based control, Model Predictive Control (MPC) has emerged as a prominent optimization-based framework for hybrid energy systems. By explicitly accounting for system dynamics, operational constraints, and forecasts of future disturbances, MPC can coordinate continuous control actions and categorical decisions in a predictive manner. This flexibility, however, comes at the cost of solving a constrained optimization problem at every sampling instant. For hybrid systems, these problems naturally take the form of Mixed-Integer Nonlinear Programs (MINLPs), whose nonlinear and combinatorial structure can make them computationally demanding \cite{grossmann2002review}.

Several approaches have been proposed to reduce this computational burden. One strategy is to simplify the system model, for example through linearization, thereby transforming the problem into a Mixed-Integer Linear Program (MILP). This enables faster optimization at the expense of model mismatch \cite{6899700}. Other approaches approximate or decompose the original problem to reduce its computational complexity. The Combinatorial Integral Approximation (CIA) method \cite{a15040121}, for instance, solves a sequence of continuous and mixed-integer optimization problems to obtain feasible discrete decisions. Online computational effort can also be reduced by adopting shorter prediction horizons, longer control intervals, or hierarchical control architectures, at the expense of reduced flexibility or optimality \cite{rawlings2022model, lottis2025minlp}.

For district heating networks, integer decisions and detailed nonlinear dynamics are less commonly addressed within the same optimization framework. Most works either model the network dynamics in detail while retaining a purely continuous formulation \cite{deGiuli2024,krug2021,sibeijn2025}, or incorporate integer variables using simplified system models \cite{taylor2021,taieb2025}. Nevertheless, some of the computational strategies discussed above have been applied to district heating control, as demonstrated by \cite{jansen2024minlp}, which employs the CIA method. Despite this, repeatedly solving complex MINLPs online remains a major challenge for real-time deployment. Moreover, existing MINLP-based approaches often rely on commercial optimization solvers, which can limit their scalability and accessibility for industrial deployment \cite{lottis2025minlp,jansen2024minlp}.

The computational burden associated with repeatedly solving complex MINLPs online has motivated an alternative approach based on offline learning. Rather than solving an optimization problem at each control step, these methods learn a control policy offline from data or simulation, shifting the computational effort to a training phase. Once trained, the resulting policy can be deployed online with minimal computational overhead.

Reinforcement learning (RL) has emerged as a prominent learning-based approach for complex control problems and has been extended to hybrid action spaces combining categorical and continuous decisions. Existing approaches include parameterized action spaces, in which continuous parameters are conditioned on the selected categorical action \cite{hausknecht2016deep,xiong2018parametrized,Bester2019MultiPassQF}, as well as stochastic policy formulations that jointly generate categorical and continuous actions \cite{fan2019,pmlr-v100-neunert20a}. These methods avoid the need to directly differentiate through categorical variables by relying on value-based or policy-gradient formulations.

Despite these advances, RL methods present several limitations for complex physical systems. Model-free RL typically requires extensive environment interactions to learn effective policies, which can result in substantial training data and computational requirements. Furthermore, standard RL policies do not inherently guarantee that their outputs satisfy actuator constraints. While simple bounds can often be enforced through action squashing or clipping, more complex constraints generally require additional constraint-handling mechanisms.

Applications of RL to DHNs exist, but have mostly focused on fully continuous \cite{GONG2024107672} or fully discrete control problems \cite{stepanovic2022rl,han2023rainbow}. Hybrid formulations combining both types of decisions remain comparatively less explored.

These limitations have motivated a class of model-based learning approaches that leverage differentiable models of the system dynamics to optimize control policies through gradient-based trajectory optimization. By propagating gradients through the dynamics, these methods directly optimize policy parameters based on the effect of control decisions over entire closed-loop trajectories. Representative examples include Differentiable Predictive Control (DPC)~\cite{DRGONA202280}, Short-Horizon Actor-Critic (SHAC)~\cite{xu2022shac}, and Performance Boosting (PB)~\cite{pb,rpb}.

Extending differentiable policy optimization to complex physical systems still presents important challenges. First, gradient propagation over long closed-loop horizons can become difficult due to vanishing or exploding gradients. SHAC, for example, mitigates this issue by deliberately restricting gradient propagation to short horizons~\cite{xu2022shac}. Such a restriction is particularly relevant for systems with long-term dynamics and delayed effects, such as district heating networks, where thermal storage decisions can influence operation over several hours. Second, existing differentiable control frameworks predominantly consider continuous control inputs, since standard discrete selection and sampling operations are not differentiable.

In the machine learning literature, differentiable relaxations of discrete variables, such as the straight-through estimator and the Gumbel-Softmax reparameterization, have been studied extensively~\cite{bengio2013estimating,jang2017categorical,maddison2017concrete}. More recently, these techniques have also been explored for control applications. The mixed-integer extension of DPC proposed in~\cite{boldocky2025learning}  learns an explicit policy mapping the current state and an $N$-step preview of the future disturbances to continuous and integer decisions of a linear time-invariant system. It compares three differentiable relaxations of the integrality constraints, namely a sigmoid straight-through estimator that rounds a scalar output, a Gumbel--Softmax straight-through estimator that selects one element of a finite integer set, and a learnable rounding threshold. A closely related construction is proposed in~\cite{okamoto2026accelerating} for district cooling, where a Gumbel--Sigmoid straight-through is the central relaxation mechanism. Both works consider plant models with linear state dynamics. For nonlinear dynamics, ~\cite{boldocky2026data} instead employs deterministic threshold rounding, together with a binary-variance regularization that pushes the relaxed binaries away from the rounding threshold. In concurrent work, \cite{zheng2026uphes} applies the Gumbel--Softmax straight-through estimator to a nonlinear system for open-loop day-ahead scheduling.

These approaches share several characteristics that distinguish them from the setting considered in this work. First, the learned policies are explicit solution maps of an optimal control problem, realized either as memoryless feedback laws of the plant state ~\cite{boldocky2025learning,okamoto2026accelerating,boldocky2026data} or as open-loop schedules computed from the initial state alone \cite{zheng2026uphes}. In both cases they are static in their response to the system states and instead consume a forecast window of the exogenous signals, therefore requiring a separate prediction layer at deployment. Second, the policies are trained by backpropagation over short horizons, of a few tens of steps. In  \cite{zheng2026uphes}, backpropagation through a 24-step sequential rollout was found to degrade learning and was replaced by a parallel approximation of the dynamics. Third, the stochastic Gumbel relaxation is compared with deterministic alternatives only through the attained cost. How the injected noise shapes the discrete decisions of the deployed policy, and why it does so, remains unexamined. Finally, in all these works the policy is trained and evaluated on the same model of the plant, so its robustness to plant-model mismatch is not assessed.

\subsection{Main contributions}

In this work, we develop a hybrid control architecture that leverages differentiable categorical relaxations for end-to-end training through closed-loop trajectories, and demonstrate its application to a complex, full-scale district heating network. Our main contributions are:
\begin{itemize}
\item \textbf{Hybrid-action neural controller (HANC).} We propose HANC, a neural controller that jointly outputs categorical actions and continuous setpoints. HANC enables long-horizon, end-to-end training over full closed-loop trajectories using backpropagation through time (BPTT), while enforcing complex actuator constraints by design. At deployment, it operates as a causal feedback policy using only current measurements and internal memory, without requiring forecasts.
\item \textbf{Application to DHN control with improved performance.} The HANC controller is deployed for district heating network control under time-varying energy prices, jointly optimizing categorical switching decisions and continuous setpoints. Tested on a high-fidelity simulation model distinct from the training model, it achieves a $30\%$ operating-cost reduction over an industrial rule-based baseline despite the model mismatch. An analysis of the learned logit distributions further reveals that Gumbel noise injection improves categorical decision margins, providing insight into its role in learning robust switching policies.
\end{itemize}

The remainder of the paper is organized as follows. Section~\ref{sec:gumbel controller} introduces the proposed controller architecture and the differentiable formulation used for its training. Section~\ref{sec:NN4DHN} describes the district heating system considered and the corresponding training procedure. Section~\ref{sec:simulations} presents the simulation results, including an ablation study investigating the role of Gumbel noise in the training process.

\section{Hybrid Action Neural Controller}
\label{sec:gumbel controller}
In this section, we introduce the proposed hybrid-action neural controller. We begin by formulating the control problem. We then describe the architecture of the controller, and finally we detail the relaxation that makes it trainable by gradient descent.

\subsection{Problem formulation}
\label{sec:problem}

We study nonlinear discrete-time time-varying systems with hybrid control inputs. At time $t$, the plant state is $x_t \in \mathbb{R}^n$ and its output is $y_t\in \mathbb{R}^p$. The dynamics of the system are
\begin{align}\label{eq:system}
    x_t &= f_t(x_{t-1:0},\,u_{t-1:0},\,w_{t-1:0})\,, ~~~t = 1,2,\ldots\,, \nonumber\\
    y_t &= g_t(x_{t:0},\,u_{t:0},\,w_{t:0})\,, ~~~t = 1,2,\ldots\,,
\end{align}
where $w_t \in \mathbb{R}^r$ represents exogenous signals influencing the system. We consider exogenous signals with support $\mathcal{W}_t\subseteq \mathbb{R}^r$ following a random distribution $\mathcal{D}_t$, and initial conditions $x_0$ following a random distribution $\mathcal{X}_0$.

The control input $u_t$ collects the commands applied to the plant actuators and must lie in a feasible set
\begin{equation}\label{eq:actionspace}
    u_t \in \mathcal{U} \subseteq \mathbb{R}^{m_c} \times \Delta\,,
\end{equation}
where the real-valued component hosts the $m_c$ continuous-valued commands and $\Delta=\Delta_1\times\cdots\times\Delta_{m_d}$, with $\Delta_j=\{1,\ldots,C_j\}$, is a finite action space collecting the $m_d$ categorical ones. We make no convexity or connectedness assumption on $\mathcal{U}$. Consequently, the feasible values of the continuous control variables may depend on the categorical operating mode and can form disconnected subsets of $\mathbb{R}^{m_c}$, such as unions of intervals and isolated points. 
This formulation enables a single controller to jointly generate heterogeneous control commands, including independent continuous and categorical variables as well as hybrid commands in which continuous and categorical variables are coupled through the feasible action set.

To generate the inputs, we consider nonlinear, causal and time-varying control policies that react to the system output as well as to any contextual information $z_t \in \mathbb{R}^q$, drawn from a distribution $\mathcal{Z}_t$, that could improve performance. The control actions are computed as
\begin{equation*}\label{eq:policy}
    u_t = K_t\!\left(y_{t-1:0},\,z_{t:0}\right)\,,
    \quad t = 1,2,\ldots,
\end{equation*}
where $K = \{K_t\}_{t\ge 1}$ is a causal, time-varying control policy. Note that the controller $K$ can be a dynamical system, as $K_t$ can depend on the entire past history of the system.

Our goal is to synthesize a policy $K$ minimizing a given cost over a finite time horizon $T$, jointly producing the continuous and discrete actions. As optimizing directly over the policy is an infinite-dimensional problem, we restrict attention to policies $K_\theta$ parametrized by a finite number of parameters $\theta \in \mathbb{R}^{n_\theta}$, keeping the problem tractable. The optimization problem to solve is:

\begin{problem}
\label{prob:boosting} Find $\theta \in \mathbb{R}^{n_\theta}$ solving the finite-horizon Nonlinear Optimal Control (NOC) problem:
\begin{alignat}{3}
\label{NOC:tes_train}
&\min_{\theta\in\mathbb{R}^{n_\theta}} && \quad
\mathbb{E}_{\substack{x_0\sim\mathcal{X}_0\\[1pt] w_{T:0}\sim\mathcal{D}\\[1pt] z_{T:0}\sim\mathcal{Z}}}
\!\left[\,L\!\left(y_{T:0},\,u_{T:0},\,w_{T:0},\,z_{T:0}\right)\right]\\
&\operatorname{s.t.}~~ &&
    x_t = f_t\!\left(x_{t-1:0},\,u_{t-1:0},\,w_{t-1:0}\right)\,, \nonumber\\
&~~&&
    y_t = g_t\!\left(x_{t:0},\,u_{t:0},\,w_{t:0}\right)\,, \nonumber\\
&~~&&
    u_t = K_{\theta,t}\!\left(y_{t-1:0},\,z_{t:0}\right)\,, \nonumber \\
&~~&&
    u_{t} \in \mathcal{U}\,,
    ~~\forall t = 1,\ldots,T\, . \nonumber
\end{alignat}
where $L(\cdot)$ is any lower-bounded, piecewise-differentiable loss over the realized output $y_{T:0}$, the control inputs $u_{T:0}$, the disturbance $w_{T:0}$ and the contextual signals $z_{T:0}$. The expectations remove the effect of the random initial condition $x_0$, the disturbance $w_{T:0}$ and the context $z_{T:0}$ on the realized loss. The flexibility in choosing $L(\cdot)$ allows multiple objectives to be addressed jointly, including output constraints.
\end{problem}

\begin{remark}
We refer to any additional measured variable that can help improve the control action as \textit{context}. We focus on settings in which the contextual information $z$ not only informs the controller but also directly affects its performance through the loss $L$. This generic formulation captures policies that must react to external signals that are informative of performance. Such signals are common in energy systems, with dynamic electricity prices or meteorological data being typical examples, and should therefore be accounted for both as policy inputs and in the cost when optimizing controllers for these applications.
\end{remark}


\subsection{Controller architecture}
\label{sec:control-architecture}

\begin{figure}
  \centering
  \includegraphics[width=\linewidth]{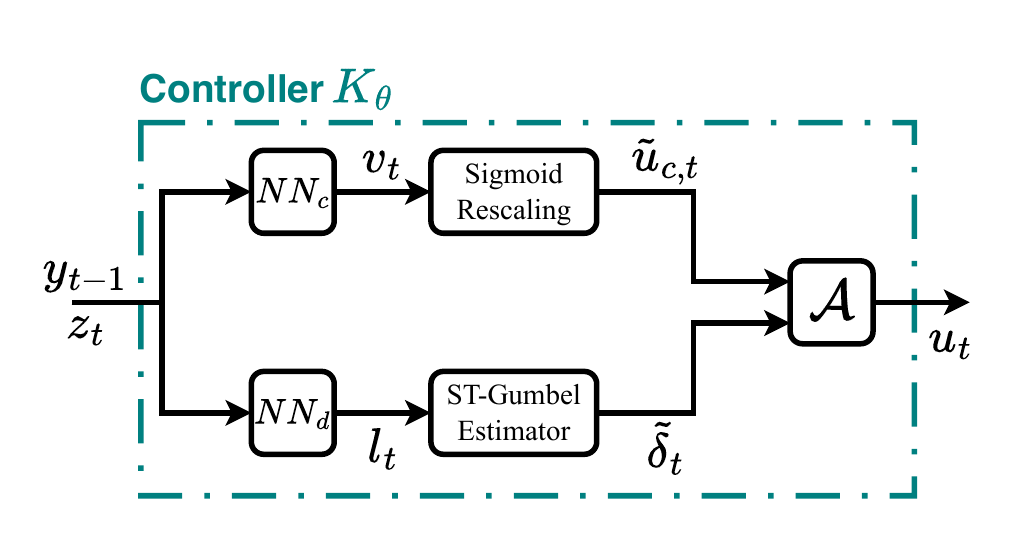}
  \caption{Generic layout of the proposed HANC architecture. The context and outputs are processed by two recurrent networks, whose outputs are shaped by a sigmoid rescaling and a categorical selection layer respectively, and composed by the assembly layer $\mathcal{A}$ into control actions satisfying actuator constraints. The categorical selection layer is implemented as a Straight-Through Gumbel estimator during training, as detailed in Section~\ref{sec:training}.}
  \label{fig:K architecture}
\end{figure}

Problem~\ref{prob:boosting} is a nonlinear optimal control problem with mixed integer decision variables. We parameterize the policy $K_\theta$ by neural networks, which provide the expressivity required to represent a rich class of nonlinear hybrid feedback laws while maintaining low computational cost at deployment. To generate these control actions through a single differentiable map, we propose the \emph{HANC} architecture, a neural architecture comprising a continuous branch that generates real-valued internal control variables $\tilde u_c$, a discrete branch that generates categorical internal control variables $\tilde\delta$, and a differentiable assembly layer $\mathcal{A}$ that maps these internal variables to the plant actuator inputs $u$. Together, these elements realize the policy $K_\theta$ while guaranteeing $u_t\in\mathcal{U}$ by design. The architecture is illustrated in Fig.~\ref{fig:K architecture}, and its components are described next.

\subsubsection*{Continuous branch}

Continuous internal control variables are generated by a neural network $NN_c$. These variables are real-valued and the branch is designed to be differentiable by construction. Common input constraints can nevertheless be enforced by appropriately designing the output layer. For box constraints with lower and upper bounds $u_{\min}$ and $u_{\max}$, the continuous control variables are computed as
\begin{equation}
\label{eq}
\tilde u_{c,t}=\sigma(v_t)(u_{\max}-u_{\min})+u_{\min},
\end{equation}
where $v_t$ denotes the output of $NN_c$ and $\sigma$ is the sigmoid function. This guarantees $\tilde u_{c,t}\in[u_{\min},u_{\max}]$ by construction.

\subsubsection*{Discrete branch}
The discrete branch neural network $NN_d$ outputs a logit vector $l_{t}$ parameterizing a categorical distribution over the admissible categories. Consider a categorical action $j$ with admissible values collected in the vector $\delta^{j}=[\delta^j_1,\delta^j_2, \ldots,\delta^j_{C_j}]$. The selection layer returns the hard one-hot encoding $\tilde{s}^j_t\in\{0,1\}^{C_j}$ of the selected category, with exactly one entry equal to one and all others equal to zero. The corresponding internal categorical action is recovered as
\begin{equation}
\label{eq:hard-selection}
\tilde{s}^j_t
=
\mathrm{onehot}\!\left(
\arg\max_{i\in\{1,\ldots,C_j\}}l^j_{i,t}
\right),
\qquad
\tilde{\delta}_t^{j}
=
\delta^{j\top}\tilde{s}^j_t,
\end{equation}
where $\mathrm{onehot}(i)$ denotes the $C_j$-dimensional one-hot vector associated with category $i$, with entry $1$ at the selected category and $0$ elsewhere. This guarantees that the resulting internal categorical action is a valid element of $\Delta_j$.

\begin{remark}
\label{sec:probability}
Note that taking the $\arg\max$ over the logits is equivalent to selecting the most likely category. Indeed, the logits could be mapped to categorical probabilities through the softmax transformation, which preserves their ordering.

\end{remark}

\subsubsection*{Assembly Layer}
The internal outputs $\tilde u_c$ and $\tilde\delta$ of the continuous and discrete branches can be combined by the assembly mapping $\mathcal{A}$ to generate the plant input
\begin{equation}
\label{eq}
u_t
=
\mathcal{A}\!\left(\tilde u_{c,t},\tilde\delta_t\right).
\end{equation}
The assembly layer $\mathcal{A}$ maps the internal actions to feasible plant inputs according to the actuator constraints and any dependencies between control variables. 

We now consider several notable subcases of the assembly layer that will be used in the subsequent case study. First, for purely categorical or purely continuous actuators, the corresponding branch output can be directly fed to the system, such that $\mathcal{A}$ reduces to the identity mapping.

The assembly layer becomes essential when the plant inputs couple internal continuous and categorical decisions. Consider an action whose feasible set is a finite union of intervals,
\begin{equation}
\label{eq:union-constraint}
\mathcal{U}
    =
    \bigcup_{i=1}^{C_j}\mathcal{U}_i,
    \qquad
    \mathcal{U}_i=[u_{\min}^{i},u_{\max}^{i}],
\end{equation}
where singletons ($u_{\min}^{i}=u_{\max}^{i}$) are admitted and can represent idle modes. To generate such input, the discrete branch generates an internal categorical action $\tilde\delta_t^j\in\{1,\ldots,C_j\}$ selecting the active mode, while the continuous branch generates one physically scaled internal setpoint $\tilde u_{c,t}^i\in\mathcal{U}i$ for each mode. Using the corresponding one-hot representation $\tilde{s}_t^j$, the assembly mapping can be designed as:
\begin{equation}
\label{eq:branch-mixture}
u_t
=
\mathcal{A}
\left(
\tilde{s}_t^j,
\{\tilde u_{c,t}^i\}_{i=1}^{C_j}
\right)
=
\sum_{i=1}^{C_j}
\tilde{s}_{i,t}^j\tilde u_{c,t}^i.
\end{equation}
Because the selection layer produces a hard one-hot representation in the forward pass, exactly one mode is selected at each time step. Since each $\tilde u_{c,t}^i$ is already constrained to its corresponding feasible interval $\mathcal{U}_i$, the assembled action satisfies $u_t\in\mathcal{U}$ by construction, both during training rollouts and at deployment.

The assembly mapping can also encode logical dependencies between internal actions. For example, suppose two binary internal categorical actions $\tilde\delta_t^j,\tilde\delta_t^{j'}\in \{0,1\}$ determine whether a continuous internal action $\tilde u_{c,t}^j$ is applied. The requirement that the corresponding action can only be active when both categorical conditions are satisfied can be implemented as:

\begin{equation}
\label{eq:and-gate}
    u^j_{t}
    =
    \mathcal{A}(\tilde\delta^j_t,\tilde\delta^{j'}_t,\tilde u^j_{c,t})
    =
    \left(\tilde\delta^j_t\,\tilde\delta^{j'}_t\right)\tilde u^j_{c,t},
\end{equation}

The product implements an AND gate, thereby ensuring the executed plant input satisfies the logical constraint exactly in the forward pass. More general logical and hybrid constraints can likewise be encoded through the assembly mapping.

\subsection{Differentiable training of the controller}
\label{sec:training}

The architecture of Section~\ref{sec:control-architecture} defines a well-posed feedback policy, but it is not yet amenable to gradient-based optimization. Were the control actions purely continuous, Problem~\ref{prob:boosting} could be addressed with established gradient-based tools, since a finite parametrization $\theta$ and a differentiable closed loop allow the expected cost to be minimized by backpropagation through time along the rollout trajectories. The selection layer \eqref{eq:hard-selection} precludes this. The $\arg\max$ is piecewise constant, hence it has zero gradient almost everywhere and an undefined gradient at transitions, which breaks the gradient chain that BPTT \cite{58337} relies on. The Gumbel-Softmax reparameterization, reviewed next, restores a usable gradient path through the discrete actions and enables end-to-end training of the full hybrid policy.

\subsubsection{Gumbel reparameterization trick}
\label{sec:gumbel}

We briefly review the Gumbel-Softmax reparameterization \cite{jang2017categorical}, which enables differentiable sampling from categorical distributions. This is done by building on the reparameterization trick in \cite{kingma2014autoencoding}, which expresses a random sample as a deterministic function of the distribution's parameters and an independent, parameter-free noise source.

Given a categorical distribution $\mathcal{C}$ over $C$ categories ${1,\ldots,C}$ with logits $l=(l_1, \ldots,l_C)$ parameterizing its probability distribution, the one-hot encoding $s \in \{0,1\}^C$ of a sample can be obtained using independent Gumbel noise $g_i$ as:
\begin{equation}
\begin{aligned}
\label{eq:argmax}
    s &= \mathrm{onehot}\!\left( \arg\max_{i \in \{1,\dots,C\}} \big( l_i + g_i \big) \right), \\
    g_i &= -\log(-\log u_i), \quad u_i \overset{\text{i.i.d.}}{\sim} \mathrm{Uniform}(0,1).
\end{aligned}
\end{equation}

The outcome of \eqref{eq:argmax} is equivalent to sampling from the categorical distribution parametrized by $l$.

However, the $\arg\max$ operation is non-differentiable and therefore prevents gradient-based training. To make the sampling operation differentiable, the Gumbel-Softmax relaxation replaces the $\arg\max$ operation with a softmax, while retaining the same Gumbel perturbations. This produces a continuous and differentiable relaxation $\hat{s}$ that approaches the one-hot sample $s$ as the relaxation parameter $\tau >0$ goes to zero:

\begin{equation}
    \label{eq:softmax}
    \hat{s}_i = \frac{\exp\!\big((l_i + g_i)/\tau\big)}{\sum_{j=1}^{C}\exp\!\big((l_j + g_j)/\tau\big)},
    \qquad i = 1, \dots, C.
\end{equation}

Thanks to the soft relaxation, all operations remain differentiable, allowing gradients to propagate from the loss through the sampled actions to the neural network parameters. The relaxation thus trades fidelity to the discrete sampling operation for differentiability.

\subsubsection{Straight-Through Gumbel estimator}
\label{sec:st-gumbel}

The relaxation \eqref{eq:softmax} is differentiable but real-valued, so using it directly in the forward pass would drive the plant with fractional operating modes that no actuator can realize. To keep the executed actions discrete while restoring a gradient path, we employ the Straight-Through (ST) Gumbel estimator introduced in \cite{jang2017categorical}, which induces different forward and backward passes.

During the forward pass, the deterministic selection \eqref{eq:hard-selection} is replaced by the Gumbel-max sampling \eqref{eq:argmax}, which returns a hard one-hot encoding $s^j_t$ and hence an executed categorical component
\begin{equation*}
    \delta_t^{j}=\delta^{j\top}s^j_t\,.
\end{equation*}
The selection remains hard, so the closed-loop system is simulated using true discrete control actions throughout training, exactly as at deployment.

\begin{figure}[t]
    \centering
    \includegraphics[width=\linewidth]{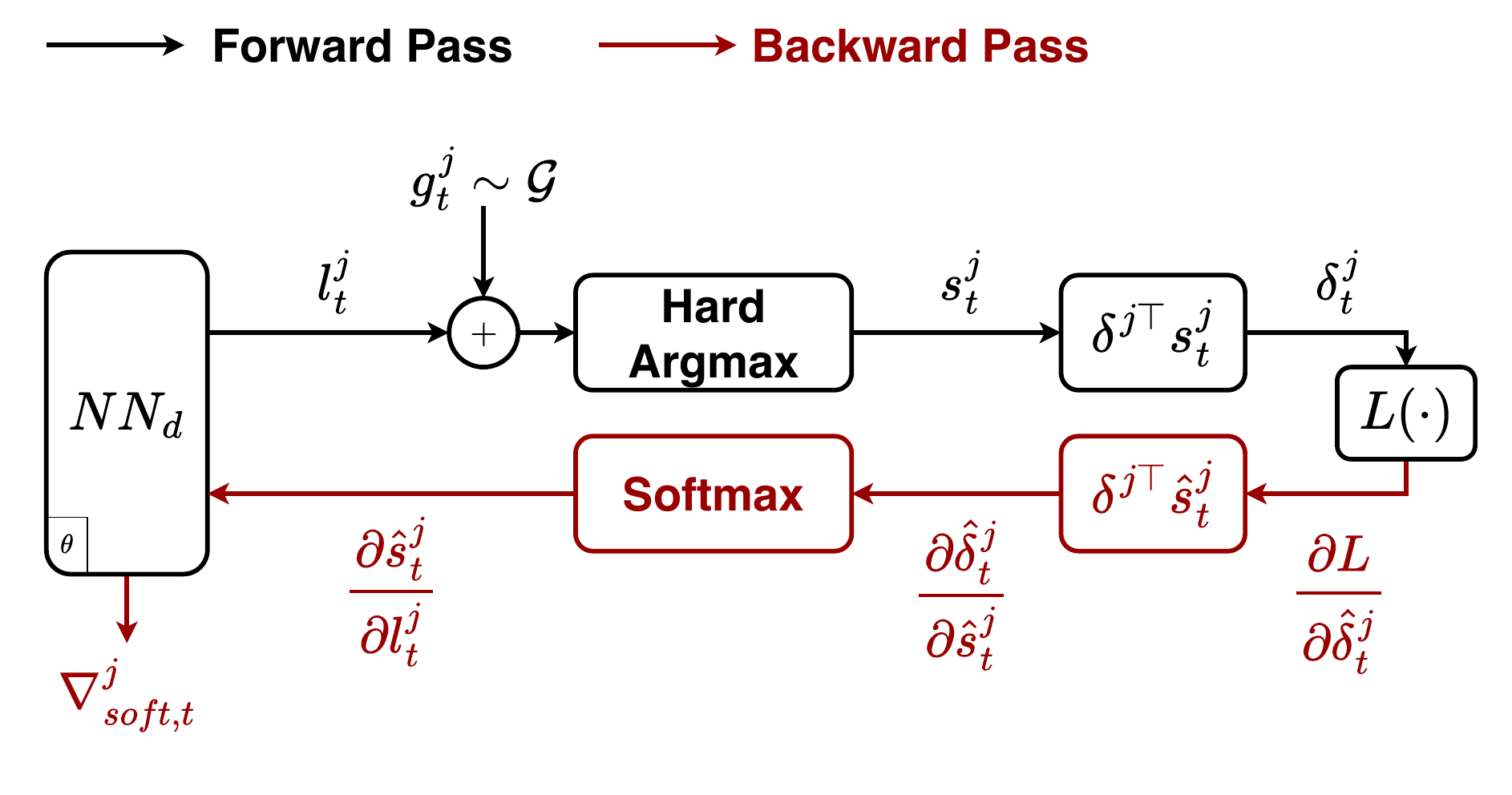}
    \caption{ST Gumbel estimator for a discrete control action $\delta^j$. In the forward pass (black), Gumbel noise $g_t^j$ perturbs the logits $l_t^j$, and the hard one-hot sample $s_t^j$ of~\eqref{eq:argmax} selects the executed action $\delta_t^j$. In the backward pass (red), gradients flow through the relaxation $\hat{s}_t^j$ of~\eqref{eq:softmax}, with the Jacobian of each block reported below the corresponding arrow.}
    \label{fig:Gumbel flow}
\end{figure}

During backpropagation, however, the gradients are computed through the relaxed action
\begin{equation*}
    \hat\delta_t^{j}=\delta^{j\top}\hat s^j_t\,,
\end{equation*}
obtained by using the relaxed sample $\hat{s}_t^{j}$ from~\eqref{eq:softmax} instead of the hard one-hot encoding. The backward pass thus remains fully differentiable, allowing the controller parameters to be optimized using standard BPTT. The forward and backward passes through the selection layer are shown in Figure~\ref{fig:Gumbel flow}.

The same mechanism propagates through the assembly mappings. In \eqref{eq:branch-mixture}, the relaxed weights $\hat{s}^j_{i,t}$ distribute the gradient across all candidate setpoints during backpropagation, providing each operating mode with a learning signal even though only one is executed in the forward pass.

\subsubsection{Annealing}
\label{sec:annealing}


As $\tau \to 0$, the softmax relaxation converges in distribution to the underlying discrete distribution, so the mismatch between relaxed and true categorical distribution vanishes. However, this regime is associated with increasingly ill-conditioned gradients, which can hinder numerical optimization. To address this issue, \cite{jang2017categorical} propose an annealing strategy in which $\tau$ is initialized at a high value and gradually decreased during training. This allows the model to benefit from smoother gradients in the early stages of optimization, facilitating convergence to a suitable region of the parameter space, before progressively transitioning toward a more accurate approximation of the discrete distribution. 

In this work, we use an exponential annealing schedule over the training epochs $k$: 

\begin{equation}
\label{eq:tau-schedule}
    \tau_k=\max\!\left(\tau_0\,\rho^{k},\,\tau_{\min}\right),
    \qquad \rho\in(0,1)\,.
\end{equation}
With this approach, early training benefits from smooth, well-conditioned gradients while late training operates close to the discrete regime. The floor $\tau_{\min}>0$ preserves gradient flow throughout. As we utilize a ST estimator, the categorical inputs constraints are guaranteed during the forward pass and at deployment.

\subsubsection{Stochastic exploration and decision margins}

Injecting Gumbel noise makes the controller stochastic during training. This improves optimization in two ways. First, it promotes exploration by perturbing discrete decisions and helps escape poor local minima. Second, the controller is rewarded for producing logits whose sampled decisions remain robust to these perturbations, encouraging confident decisions away from switching boundaries. These optimization benefits have also been observed empirically in \cite{jang2017categorical}, and are further explored in Section~\ref{sec:results-gumbel-vs-ste}. At deployment, the Gumbel noise is disabled and the selection layer reverts to the deterministic $\arg\max$ of \eqref{eq:hard-selection}, so the implemented controller remains a deterministic causal feedback policy.

The proposed architecture and training setup therefore realizes the controller $K_\theta$ of Problem~\ref{prob:boosting} while generating hybrid control actions satisfying prescribed input constraints by design. Since these constraints are enforced through the controller architecture rather than through additional optimization constraints, Problem~\ref{prob:boosting} reduces to an unconstrained optimization over $\theta$. Moreover, as the computational burden is incurred entirely during offline training, deployment only requires a single forward pass through the neural networks, making the controller computationally inexpensive in real-time operation.

\section{Neural Network control of a DHN}
\label{sec:NN4DHN}

The HANC framework introduced above is intended for control problems in which complex dynamical systems must be operated through a combination of discrete and continuous actions. District heating networks provide an instance of this class of problems, as their multiple interacting components, thermal dynamics, operational constraints, and discrete operating decisions result in a challenging mixed-action control problem. In this section, we therefore demonstrate the application of the HANC framework to a DHN.

We first describe the DHN system considered and its control problem. We then introduce the differentiable discrete-time surrogate model $\phi_t,\gamma_t$, which approximates the input-output dynamics of the true plant and enables backpropagation through time. Finally, we detail how the HANC is trained through closed-loop rollouts of the differentiable surrogate.

\subsection{System considered}
\label{sec:dhn-system}

\begin{figure*}[t]
    \centering
    \includegraphics[width=0.98\textwidth]{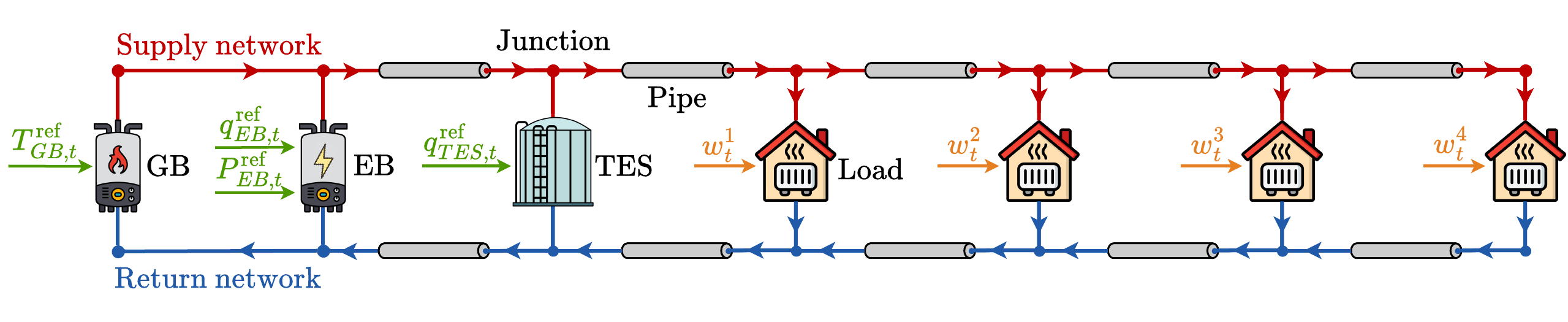}
    \caption{Schematic layout of the district heating network considered in this
work. Red and blue arrows indicate the nominal flow directions in the supply
and return networks, respectively. Green arrows denote the control variables,
whereas orange arrows denote the exogenous thermal-demand disturbances.}
    \label{fig:dhn-layout}
\end{figure*}

District heating networks distribute thermal energy from one or more
generation units to spatially distributed consumers through a closed-loop
water network composed of supply and return lines. The layout of the district
heating network considered in this work is shown in
Fig.~\ref{fig:dhn-layout}.

The system considered in this work is based on the district heating network
benchmark located at the RSE S.p.A. Thermal Facility~\cite{nigro2025experimental}.
A validated dynamic simulator of the network was developed in the OpenModelica
environment based on the physical modeling framework of~\cite{nigro2025}
and the DHN4Control library~\cite{alvarado2024development}. This Modelica
simulator is adopted throughout this work as the reference system for
surrogate-model identification and for the closed-loop evaluation of the
trained controller. Its layout is shown in Fig.~\ref{fig:dhn-layout}.

The plant comprises a gas boiler (GB), an electric boiler (EB), a stratified
thermal energy storage (TES) unit, and four thermal loads interconnected by the
distribution network. The GB is operated through its controllable supply temperature reference $T^{\mathrm{ref}}_{GB,t}$ and acts as the slack heat generator of the network. From a hydraulic perspective, it receives the portion of the total return mass flow that does not flow through the EB or TES branches. Consequently, its mass flow rate is not independently controlled, but results from the network flow balance as the residual return flow.

The EB is equipped with an internal circulation pump with controllable mass
flow rate $q^{\mathrm{ref}}_{EB,t}$ and controllable supplied thermal power $P^{\mathrm{ref}}_{EB,t}$. Owing
to the admissible operating conditions of the internal pump and of the
electrical heating unit, both quantities are subject to hybrid operating
constraints. In particular, the EB can either be switched off or operate
within its admissible active range, such that

\[q^{\mathrm{ref}}_{EB,t}\in\{0\}\cup[\,q_{EB,\min},\,q_{EB,\max}\,]\]
and
\[P^{\mathrm{ref}}_{EB,t}\in\{0\}\cup[\,P_{EB,\min},\,P_{EB,\max}\,]\]
The zero value corresponds to the off state, while values inside the
corresponding intervals represent admissible operating conditions when the EB
is active.

The thermal demand of load $\ell$ at time $t$ is denoted by $w_t^\ell$,
$\ell=1,\ldots,4$, and the vector
\[ w_t = \left(w_t^1,\ldots,w_t^4\right)\]
collects the thermal demands acting on the network.

The generation units, i.e. GB, EB and TES draw water from the return network, increase its
temperature, and inject it into the supply network. The heated water is
transported through the supply pipelines toward the thermal loads, where
thermal energy is extracted to satisfy the local heat demand. The resulting
colder water is then conveyed back to the generation units through the return
network, closing the hydraulic circuit. At each load, the delivered thermal
power depends on the water mass flow rate and on the temperature difference
between the supply and return sides.

The thermohydraulic behavior of the network is determined by the interaction
among water temperatures, mass flow rates, and the dynamics of its components.
Distribution pipes introduce thermal losses and transport delays, so
variations in generation or demand propagate through the network only after a
finite time. At hydraulic junctions, water streams are split or mixed
according to mass and energy conservation. Consequently, changes in the
operating conditions of one component affect the temperatures and mass flow
rates at other locations in the network, with these effects propagating
according to the hydraulic topology and transport dynamics.

The TES provides additional operational flexibility by partially decoupling
heat generation from instantaneous demand. It is connected to the network
through a bidirectional pump with controllable mass flow rate $q^{\mathrm{ref}}_{TES,t}$.
Depending on the flow direction, the TES can either absorb thermal energy from
the network during charging or release previously stored energy during
discharging.

Due to the operating constraints of the bidirectional pump, active operation
requires a minimum differential pressure, corresponding to a minimum
admissible mass-flow magnitude $q_{TES,\min}$, while the pump also imposes
a maximum admissible mass-flow magnitude $q_{TES,\max}$. The TES can additionally
remain idle, in which case no water is exchanged with the network. Its
admissible mass-flow set is therefore

\[
\begin{aligned}
q^{\mathrm{ref}}_{TES,t} \in {}&
[-q_{TES,\max},-q_{TES,\min}]
\cup\{0\} \\
&\cup [q_{TES,\min},q_{TES,\max}].
\end{aligned}
\]

The two nonzero intervals correspond to the two possible flow directions
through the storage connection, while $q^{\mathrm{ref}}_{TES,t}=0$ represents the idle state.

Owing to its stratified structure, the TES stores water at different
temperatures across its height. Its thermal state is characterized through four
temperature measurements $T_{s,t}^{i}$, $i=1,\ldots,4$, corresponding to
equally spaced locations along the tank height. The state of the storage
therefore depends not only on the total stored energy but also on its internal
temperature distribution. This long-term thermal memory enables heat production
to be shifted over time, allowing the generation units to be coordinated
according to both thermal demand and energy prices.

\subsection{System modeling}
\label{sec:model}
Closed-loop training of the HANC requires a differentiable surrogate of the reference system dynamics. To improve the physical consistency of the trajectories generated during controller optimization, we adopt the topology-guided physics-informed modeling approach proposed in~\cite{sgadari_2026_topology_pil}. The method decomposes the district heating network into interconnected component models following the physical network topology, while enforcing mass and energy conservation laws directly through the model architecture. This provides a fully differentiable surrogate while preventing violations of the main network conservation principles by construction. The general modeling and identification framework is described in detail in~\cite{sgadari_2026_topology_pil}.
For the network considered here, the dynamics of the four thermal loads and eight distribution pipes are represented by individual GRU-based submodels. Each GRU has three latent states and a single recurrent layer. Conversely, components for which sufficiently informative physical relationships are available are modeled directly: the TES is represented through a four-layer stratified thermal model according to~\cite{nigro2025}, while the EB outlet temperature is computed from its inlet temperature, supplied thermal power, and mass flow rate. Hydraulic flow balances and temperature mixing at network junctions are imposed analytically according to the network topology, following the physics-informed interconnection structure of~\cite{sgadari_2026_topology_pil}.

The identification dataset is generated by simulating the validated Modelica reference system described in Section~\ref{sec:dhn-system}, with a sampling time of $5$ min over $20$ days. The controllable inputs are excited using pseudorandom multilevel signals spanning their admissible operating ranges, while synthetic thermal-demand profiles are generated to reproduce the characteristic temporal behavior of the measured load demands. After discarding the initial transient, the dataset is split chronologically into $70\%$, $15\%$, and $15\%$ subsets for training, validation, and testing, respectively.

The model parameters are optimized using Adam with a learning rate of $10^{-3}$ for $500$ epochs, minimizing the mean-square prediction error on normalized variables. Training is performed on overlapping sequences of $64$ samples. The resulting surrogate contains $820$ trainable parameters. On the independent normalized test dataset, the model achieves an RMSE of $3.75\times 10^{-2}$ and a FIT index of $87.94\%$ ($R^2=0.886$). The identified model is then used as the differentiable system model for the closed-loop controller training described in Section~\ref{sec:training}.

\subsection{Control design and training
}

We now look at how the controller $K_\theta$ of
Section~\ref{sec:control-architecture} is trained by closed-loop rollout
through the differentiable surrogate. 

First, we reformulate Problem~\ref{prob:boosting} in a tractable way. We therefore draw a training set of $N$ scenarios
{\small
\begin{equation*}
\mathcal{S}
=
\left\{\left(x_0^{(i)},w_{T:0}^{(i)},z_{T:0}^{(i)}\right)\right\}_{i=1}^{N},
\quad
x_0^{(i)}\!\sim\mathcal{X}_0,\,
w^{(i)}\!\sim\mathcal{D},\,
z^{(i)}\!\sim\mathcal{Z}
\end{equation*}
}
and replace the expected cost with its empirical counterpart, yielding
the sampled version of Problem~\ref{prob:boosting}:

\begin{problem}
Empirical NOC
\label{prob:tractable}    

\begin{alignat}{2}
\label{eq:empirical-NOC}
    &\min_{\theta\in\mathbb{R}^n_\theta}
    &&\quad
    \frac{1}{N}\sum_{i=1}^{N}
    L\!\left(\hat y^{(i)}_{T:0},\,u^{(i)}_{T:0},\,\,
             w^{(i)}_{T:0},\,z^{(i)}_{T:0}\right)\\
    &\operatorname{s.t.}
    &&\quad
    h^{(i)}_t=\phi_t\big(h^{(i)}_{t-1:0},\,u^{(i)}_{t-1:0},\,
                  \,w^{(i)}_{t-1:0}\big)\,,\nonumber\\
    &&&\quad
    \hat y^{(i)}_t=\gamma_t\big(h^{(i)}_{t:0},\,u^{(i)}_{t:0},\,
                    \,w^{(i)}_{t:0}\big)\,,\nonumber\\
    &&&\quad
    u^{(i)}_{t}
        =K_{\theta,t}\big(\hat y^{(i)}_{t-1:0},\,z^{(i)}_{t:0}\big)\,,
    \qquad \forall t,\forall i .\nonumber
\end{alignat}
\end{problem}

where the true dynamics $(f_t,g_t)$ of \eqref{eq:system} have been
replaced by the differentiable surrogate $(\phi_t,\gamma_t)$ over latent state $h_t$. The context $z$ represents all the measurable information helping in controlling the DHN, and will be explicitly defined in Section \ref{sec:signals}.
Every operation appearing in \eqref{eq:empirical-NOC} is differentiable in $\theta$, and the input constraints are
enforced by the controller architecture. Problem~\ref{prob:tractable} is therefore an
\emph{unconstrained} stochastic program: at every epoch a mini-batch of
scenarios is sampled from $\mathcal{S}$, the closed loop
$K_\theta\!\to\!\phi\!\to\!K_\theta$ is unrolled over the full
horizon $T$, and the gradient of the batch loss is obtained by
backpropagation through time.

We now look into how the problem loss is designed. The loss $L$ dictates what the controller learns; its design is
therefore as important as the architecture itself. We construct $L$ as a
weighted sum of terms that are all \emph{dimensionless and of comparable
magnitude}, so that the weights $\alpha$ directly express priorities
rather than compensate for heterogeneous units. We present each term in
turn before assembling the total loss.
 
\subsubsection*{Economic cost}
Heat generation in the DHN incurs the per-step operating cost
\begin{equation}
\label{eq:loss-perf}
c_t
=
t_{\mathrm{s}} \left(\frac{P_{GB,t}}{\eta_{GB}}\,p^{g}t
+\frac{P_{EB,t}}{\eta_{EB}}\,p^{e}t,\right),
\end{equation}
where $P_{GB,t}$ and $P_{EB,t}$ are the gas boiler and electric boiler thermal powers computed from the plant outputs, $\eta_{GB}$ and $\eta_{EB}$ are the corresponding efficiencies, $t_s$ is the sampling time, and $p^{g}_t$ and $p^{e}_t$ are the gas and electricity tariffs.
We minimize a normalized version of this cost, defined in a regret-like fashion relative to per-scenario envelopes. Let
$D_t=\sum_{\ell}w^{\ell}t$ denote the total thermal demand. The lower and upper envelopes are defined as
{\begin{equation*} \small \begin{aligned} \underline{J} &=\frac{1}{T}\sum_t  t_{\mathrm{s}} D_t \min\!\left\{\frac{p^g_t}{\eta_{GB}}, \frac{p^e_t}{\eta_{EB}}\right\},\\ \overline{J} &=\frac{1}{T}\sum_t t_{\mathrm{s}} D_t \max\!\left\{\frac{p^g_t}{\eta_{GB}}, \frac{p^e_t}{\eta_{EB}}\right\}. \end{aligned} \end{equation*}}
These correspond to serving the exogenous demand entirely with the effectively cheaper and more expensive source, respectively, at each time step. The economic loss is then
\begin{equation}
\label{eq:econ-cost}
\ell_{\mathrm{cost}}
=
\frac{\tfrac{1}{T}\sum_{t=0}^{T} c_t\;-\;\underline{J}}
{\overline{J}-\underline{J}}\,.
\end{equation}
This normalization is invariant to a common multiplicative scaling of the tariffs, yielding a dimensionless measure that is close to $0$ for a policy tracking the cheaper source and close to $1$ for one tracking the more expensive source. Crucially, the envelopes are anchored to the \emph{exogenous} demand $D_t$ rather than to the total generation. Consequently, arbitrage opportunities remain visible to the loss: cost reductions obtained by shifting energy in time, such as charging the storage with cheap heat and discharging it during expensive periods, can result in $\ell_{\mathrm{cost}}<0$.

\begin{remark}
    The naive choice would be to minimize directly this cost $L=\tfrac{1}{T}\sum_t c_t$, but this approach suffers from two defects.
First, its scale is dictated by the realized tariffs: scenarios with
high price levels dominate the batch gradient regardless of how well
the controller performs in them. Second, the fraction of the cost that
the controller can actually influence is small compared to the
incompressible cost of serving the demand, so the useful gradient
signal is weak. 
\end{remark}

\subsubsection*{Physical-consistency penalties}
Training in closed loop with a learned surrogate exposes a specific
failure mode: the optimizer will actively seek out regions of the
input space where the surrogate deviates from the true plant whenever
such deviations lower the loss, effectively exploiting model error rather than solving the control problem. Since the surrogate of
Section~\ref{sec:model} is physics-informed but not exact, we
guard against this by penalizing trajectories that are physically
inconsistent or lie outside the plant's realizable operating envelope.
Pointwise-in-time conditions, written as $h_k(y_t)\le 0$ for
$k=1,\ldots,n_h$, are handled by smooth one-sided penalties
\begin{equation*}
\label{eq:loss-con}
    \ell_{h_k}
    =
    \frac{1}{T}\sum_{t=0}^{T}
    \frac{\mathrm{softplus}\big(h_k(y_t)\big)^{2}}{\bar{h}_k^{2}}\,,
\end{equation*}
where $\bar{h}_k$ is a characteristic scale rendering the term
dimensionless, and where $\mathrm{softplus}(x) = \log(1+e^{x})$ is a smooth approximation of $\max\{0,x\}$ which penalizes excursion from the acceptable range. Furthermore, the square makes the penalty increasingly severe with the violation depth. This
template is applied to the following conditions, each of which the
learned model could otherwise be driven to violate:
\begin{itemize}
    \item \emph{Heat delivery}
    ($h = w^\ell_t - P^{\ell}_{\mathrm{del},t}$, $\bar h = w^\ell_t$):
    $P^{\ell}_{\mathrm{del},t}$ is the power actually delivered to load
    $\ell$ and $w^\ell_t$ the power it requests. In the real network a
    load always draws its requested power, so the two coincide by
    construction; the surrogate, however, can produce trajectories in
    which they differ. Under-delivery is therefore penalized as a
    regime with no physical counterpart.

    \item \emph{Storage realizability}
    ($h=T^{\mathrm{ret}}_{s,t}-T^{\mathrm{sup}}_{s,t}$,
    $\bar h=10\,^{\circ}$C): when either in charging or discharging mode, a TES return temperature
    exceeding its supply temperature is a
    stratification inversion under which no useful heat would flow in
    the physical tank; the surrogate tolerates it, so it must be
    penalized. The term is gated by $|q^{\mathrm{ref}}_{TES,t}|/(|q^{\mathrm{ref}}_{TES,t}|+\epsilon)$, so that it vanishes when
    the storage is idle. We chose $\bar h=10\,^{\circ}$C as it is the order of magnitude of the TES temperatures.
    \item \emph{Gas boiler flow direction}
    ($h=-q_{GB,t}$, $\bar h=q^{\mathrm{nom}}_{GB}=1.5$~kg/s): 
    a negative gas boiler mass flow is nonphysical and should be penalised. 
\end{itemize}

The scales $\bar h_k$ are not tuned hyperparameters. They fix the units
in which each violation is measured, so that the weights
$\alpha_{h_k}$ express relative priorities rather than compensate for
heterogeneous magnitudes. Each is set to a representative magnitude of
the quantity it normalizes, and only its order of magnitude is
relevant.

Beyond these pointwise constraints, the surrogate may also produce trajectories that are inconsistent with the overall energy balance, potentially consuming more energy than truly available in the network. To discourage such behavior, we introduce an energy balance term over the full trajectory,
\begin{equation*}
\label{eq:loss-phys}
    \ell_{\mathrm{phys}}=
    \frac{\psi^{2}}{\bar{\psi}^{2}}\,,
    \qquad
    \psi
    =
    E_{\mathrm{gen}}
    -(1+\kappa)\,E_{\mathrm{load}}
    -\Delta E_{\mathrm{TES}}\,,
\end{equation*}
where $E_{\mathrm{gen}}$ and $E_{\mathrm{load}}$ denote the generated and delivered thermal energies over the full trajectory, respectively. The parameter $\kappa=0.05$ accounts for distribution losses, while $\Delta E_{\mathrm{TES}}$ is the change in stored energy computed from the layer temperatures and $\bar{\psi}$ is a characteristic energy scale. 

\subsubsection*{Operative constraints} 
Given that we are optimizing over a precise model of the district heating network, we are also able to enforce some operating constraints on the outputs across the whole network. In particular, we aim at making the supply temperature at the loads high enough compared to the return network so that heat can flow smoothly. As the return network tracks a temperature of 60°C, we aim to keep the supply temperatures above $T_{sup}^{min}=65^\circ$C. We implement this with a one-sided penalty: 

\begin{equation*}
\label{eq:loss-con}
    \ell_{op}
    =
    \frac{1}{T}\sum_{t=0}^{T}\sum_{l=1}^{4}
    \frac{\mathrm{softplus}\big(T_{sup}^{min}-T_{sup,t}^l\big)^{2}}{{T_{sup}^{min}}^2}\,,
\end{equation*}

\subsubsection*{Switching regularization}
Relaxed discrete decisions can oscillate rapidly between categories,
a behavior that is harmless in the surrogate but translates into
actuator chattering on the real plant. We penalize the discrete
second-order difference of the \emph{relaxed} weights of the storage
categorical and of the two electric boiler
gates,
\begin{equation*}
\label{eq:loss-switch}
    \ell_{\mathrm{sw}}
    =
    \frac{1}{T-1}\sum_{t=1}^{T-1}
    \big\|\hat{s}_{t+1}-2\hat{s}_{t}+\hat{s}_{t-1}\big\|_2^{2}\,.
\end{equation*}
The penalty acts on the relaxed weights $\hat{s}_t$ rather than on the executed one-hot samples: the latter are piecewise constant in $\theta$ and would return no gradient, whereas $\hat{s}_t$ carries smooth information about how close the controller is to switching. Penalizing the second difference rather than the first discourages rapid oscillations, which is the signature of chattering, while leaving isolated, sustained mode changes almost unpenalized. A first-difference penalty would instead tax every transition and bias the policy against using the discrete actuators at
all.
 
\subsubsection*{Total loss}
The complete training loss is the weighted sum
\begin{equation*}
\label{eq:loss-total}
    L
    =
    \alpha_{\mathrm{cost}}\,\ell_{\mathrm{cost}}
    +\sum_{k=1}^{n_h}\alpha_{h_k}\,\ell_{h_k}
    +\alpha_{\mathrm{phys}}\,\ell_{\mathrm{phys}}+\alpha_{op}\, \ell_{op}
    +\alpha_{\mathrm{sw}}\,\ell_{\mathrm{sw}}\,.
\end{equation*}

Because every term is dimensionless and $O(1)$ on typical
trajectories, the weights admit a direct interpretation as relative
priorities: consistency weights should be set large enough to make
violations dominate the economic term, while the switching weight is
kept small so that it shapes, rather than overrides, the economic
behavior.

To ensure efficient training some complementary design choices are needed. First, the economic term \eqref{eq:econ-cost} produces informative gradients
only when the scenarios expose a meaningful price contrast. When
the spread between the low and high electricity levels is small, training tends to stall in minima where the
storage and switching capabilities of the plant are unused.
We address this with a curriculum on the scenario distribution.
Let $\zeta$ parametrize the distribution $\mathcal{Z}(\zeta)$ of the
tariff profiles, with $\zeta^{\mathrm{tgt}}$ corresponding to the
realistic target distribution and $\zeta^{0}$ to a variant with
exaggerated price contrast.
Training starts from $\mathcal{Z}(\zeta^{0})$, holds it for a warmup
fraction of the epochs, then interpolates linearly toward the target,
\begin{equation}
\label{eq:curriculum}
    \zeta_k
    =
    \zeta^{0}
    +\lambda_k\left(\zeta^{\mathrm{tgt}}-\zeta^{0}\right),
    \qquad
    \lambda_k=\mathrm{clip}\!\left(\tfrac{k-k_w}{K-k_w},\,0,\,1\right),
\end{equation}
where $k_w$ is the warmup length and $K$ the number of annealing epochs.
The exaggerated contrast helps identify optimal strategies and their relation with contextual signals, and drives the parameters into the correct region of the loss
landscape. The subsequent interpolation specializes the policy to the
realistic distribution within that region. A final specialization
phase holds $\zeta_k=\zeta^{\mathrm{tgt}}$ for additional epochs. The model saved is selected as the best over this specialization phase, so
that the returned controller is selected exclusively under the target
distribution.

Finally, backpropagating through hundreds of closed-loop steps compounds the
Jacobians of the surrogate and of the controller, and is well known to
produce vanishing or exploding gradients. This has motivated some related works to
consider short-horizon methods that truncate the rollout. We instead
retain the full long horizons that are essential for learning the
storage arbitrage, and stabilize training through three complementary
mechanisms. First, global gradient-norm clipping,
$\|\nabla_\theta L\|_2\le\bar{g}$, caps the magnitude of each update
while preserving the gradient direction whenever clipping is inactive.
Second, the annealing \eqref{eq:tau-schedule} maintains
smooth controller transitions during the early stages of training,
reducing the sharpness of the optimization landscape when unstable gradients are most likely to occur. Third, the
recurrent units used by the controller and by the surrogate are gated
architectures whose gating mechanisms improve gradient flow over long
sequences by mitigating vanishing and exploding gradients. Together,
these mechanisms were sufficient in our experiments to train over the
full horizon without truncation.

\section{Simulation Results}
\label{sec:simulations}

We evaluate the proposed methodology on the district heating system described in Section~\ref{sec:dhn-system}, operated under dynamic electricity pricing and incurring the operating cost defined in \eqref{eq:loss-perf}. The control objective is to minimize the total operating cost by selecting the most cost-effective heat source at each time step and exploiting the TES to shift heat production in time, while meeting the thermal demand and respecting all equipment constraints. Performance is ultimately assessed in closed loop on an independent high-fidelity Modelica plant model driven by real market prices.

\subsection{Controller signals}
\label{sec:signals}
We first introduce the signals fed to the controller $K_\theta$ and the control actions it generates. 
\label{sec:controller-instantiation}
 
\subsubsection{Measurements and context}
The controller operates on quantities that are realistically available
online. Its input at time $t$ consists of: 

\begin{itemize}
    \item The supply and return temperatures, and the massflows inside the heating station. This includes the GB, the EB and the TES. 
    \item The TES layer temperatures, available in practice from the sensor string of a stratified tank.
    \item The current tariffs $(p^{g}_t,p^{e}_t)$
    \item The current thermal demand of the four loads.
\end{itemize}

The demand acts as the disturbance $w_t$ in \eqref{eq:system}; its
current value is measurable and is therefore folded into the context
signal supplied to the policy, making the fed contextual signal $z_t = (p^{g}_t,p^{e}_t, w_t)$ .

\subsubsection{Control actions and input constraints}
The controller commands four physical actuators, each realized as an
instance of the constructions of
Section~\ref{sec:control-architecture}:
\begin{itemize}
    \item \emph{Gas boiler supply temperature}
    $T^{\mathrm{ref}}_{GB,t}\in[\,65,\,80\,]\,^{\circ}\mathrm{C}$:
    a plain box constraint, enforced by the sigmoid
    rescaling of the continuous branch.
    \item \emph{TES mass flow}
    $q_{TES,t}^\text{ref}\in
    [\,-0.5,
     \,-0.25\,]    \cup\{0\}\cup
    [\,0.25,
     \,0.5\,]$ kg/s:
    a three-interval instance of \eqref{eq:union-constraint}, realized
    by one three-way categorical (discharge/hold/charge) mixing two
    branch setpoints and the idle singleton as
    in~\eqref{eq:branch-mixture}. The categorical decision is the
    storage \emph{mode}; the continuous branch sets the magnitude
    within the active interval.
    \item \emph{Electric boiler mass flow}
    $q_{EB,t}^\text{ref}\in\{0\}\cup[\,0.5,\,1.5\,]$~kg/s:
    the binary instance of \eqref{eq:branch-mixture} with its own
    on/off gate.
    \item \emph{Electric boiler power}
    $P_{EB,t}^\text{ref}\in\{0\}\cup[\,20,\,140\,]$~kW:
    the same binary construction, additionally subject to the
    interlock that power may only be delivered while the flow is
    active. Following~\eqref{eq:and-gate}, the power gate is the
    product of the two binary gates,
    $P_{EB,t}^\text{ref}=\left(\delta^{q}_{EB,t}\,\delta^{P}_{EB,t}\right)u^P_{EB,t}$, with $\delta^{q}_{EB,t}$ and $\delta^{P}_{EB,t}$ the on/off gates of the EB massflow and power respectively, and $u^P_{EB,t}$ the continuous power control action. That way, no combination of control outputs can command power
    through a closed hydraulic circuit.
\end{itemize}
Every executed action therefore lies in the physical feasible set of
its actuator at every time step, during training and at deployment, and this is guaranteed by design. These actions are computed with a sampling time $t_S$ of 5 minutes.
 
\subsection{Dataset and training setup}
\label{sec:training-setup}
 
Each training scenario is a tuple
$\left(x_0^{(i)},w^{(i)}_{T:0},z^{(i)}_{T:0}\right)$ constructed as
follows. The load profiles $w^{(i)}$ are sampled from a synthetic
residential model producing daily profiles with morning and evening
peaks (base and peak levels of $20$ and $35$~kW per load) and
stochastic variability.
The initial TES state is drawn uniformly over
$[\,60,\,80\,]\,^{\circ}$C,
broadcast to all layers (fully mixed initial tank at a random charge
level); the remaining initial plant temperatures are fixed to nominal
operating values.
The electricity price profiles are piecewise-constant two-level
signals alternating between a low and a high tariff. The low tariff represents periods of high solar generation and consequently lower electricity prices. The two levels are drawn per scenario from
$[\,0,\,0.2\,]$€/kWh and $[\,0.4,\,1.0\,]$€/kWh respectively. The gas price is constant within each horizon and drawn uniformly between that scenario's two electricity levels.

The dataset comprises $N=360$ scenarios over a horizon of
$T=576$ steps of $5$ minutes ($48$ hours).
 
Training follows the procedure described in Section~\ref{sec:training}, using mini-batches of $36$ scenarios, the Adam optimizer with a learning rate of $10^{-3}$, and gradient-norm clipping at $\bar g=100$.
The annealing follows \eqref{eq:tau-schedule} with
$\tau_0=2$, $\rho=0.99$ and $\tau_{\min}=0.1$.
The curriculum \eqref{eq:curriculum} acts on the range of the upper
electricity price level, annealed from $\zeta^{0}=[\,7,\,13\,]$€/kWh to
$\zeta^{\mathrm{tgt}}=[\,0.4,\,1.0\,]$€/kWh. The warmup phase lasts for the first 400 epochs, followed by a 300-epoch transition and 500 final specialization epochs at the target distribution.
Both the $NN_d$ and $NN_c$ branches of the HANC use GRUs with $64$ hidden units.

For validation, the trained controller is evaluated on a high-fidelity Modelica model of the district heating network, calibrated against measurements from a real district heating network \cite{nigro2025}. Furthermore, the evaluation scenarios use real energy prices from the Italian day-ahead market, published by the market operator GME~\cite{gme_pun}.

\subsection{Results}

We first study the qualitative behavior of the HANC on a three day rollout. We then compare it quantitatively to industry standards, before analysing the value of noise injection during training by comparing Gumbel with deterministic straight through estimators.

\subsubsection{Qualitative behavior of the HANC}

\begin{figure}[h]
  \centering
  \includegraphics[width=\columnwidth]{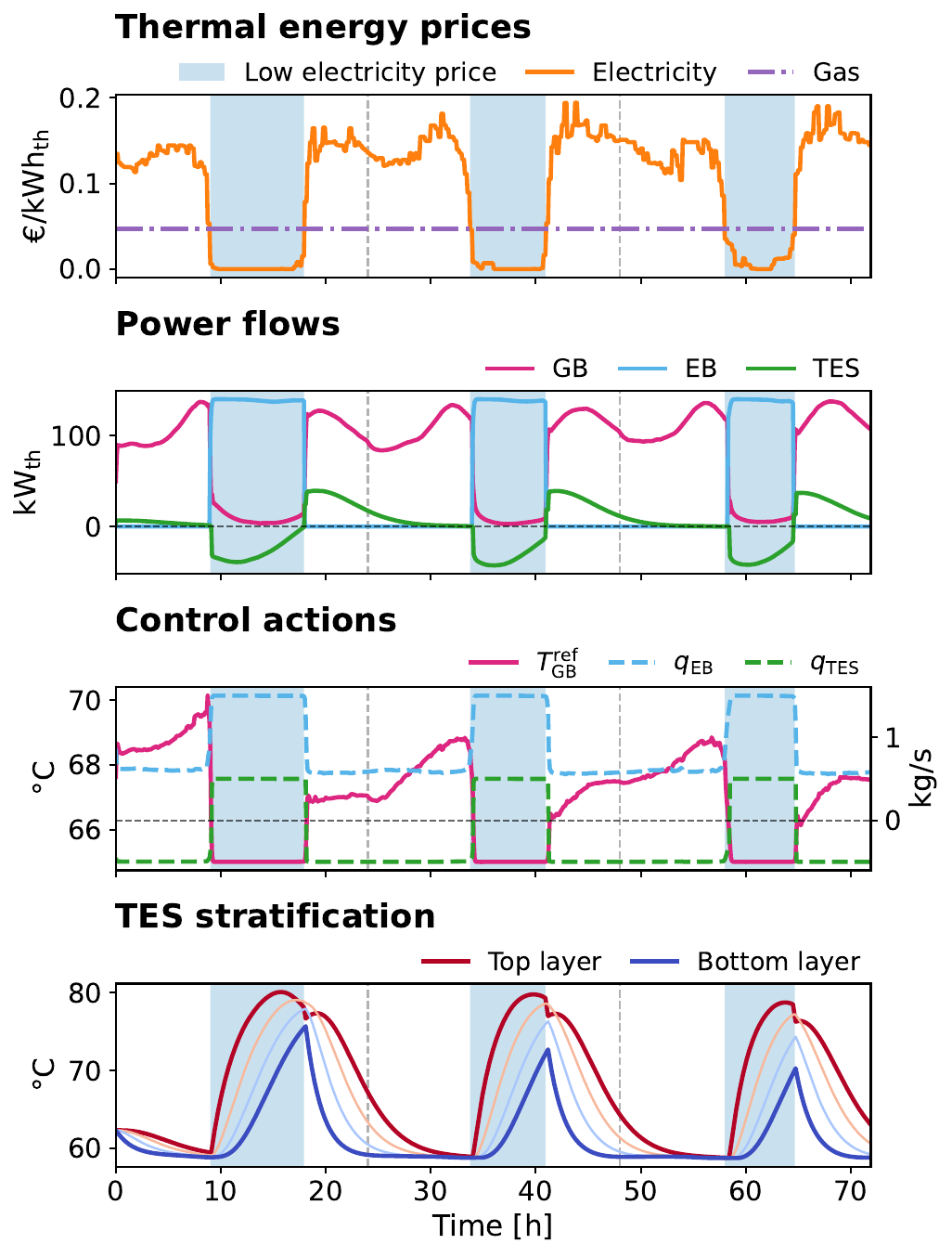}
  \caption{Closed-loop rollout of the HANC on the simulator under the real Italian price profile, shown over three representative days. The top panel gives the thermal-equivalent electricity price together with the thermal-equivalent gas price. Shaded windows mark the periods where the EB is the cheaper energy source. The second panel shows the thermal power exchanged by the GB, EB, and TES. The electric boiler covers the loads whenever electricity is cheap, and the TES switches from charging (negative values) to discharging (positive values) as the price regime changes. The third panel gives the corresponding control actions, namely the GB supply temperature setpoint (left Y-axis) and the EB and TES mass flow references (right Y-axis). The bottom panel shows the resulting stratified TES temperature profile. }
  \label{fig:trueprice_rollout}
\end{figure}

We first examine the closed-loop behavior of the HANC when deployed on the high-fidelity simulator under the real Italian price profile. Figure~\ref{fig:trueprice_rollout} shows a three-day rollout starting from a discharged TES. The top panel displays the thermal cost of energy from both generation sources. The collapse of electricity prices around midday, driven by photovoltaic generation, creates clear daily windows during which the electric boiler is cheaper than the gas boiler. In this setting the qualitatively optimal strategy is intuitive. The EB should cover the load and charge the storage during these windows, while the stored energy and the gas boiler should serve the remaining hours.

The thermal output and control action panels show that the controller reproduces exactly this strategy. The switches between generation sources occur at the shifts between pricing regimes, and the TES is charged during the cheap-electricity windows and discharged during the expensive periods. 
The stratification panel confirms this reading, with the tank temperatures tracing clean charge and discharge cycles aligned with the price periods and spanning nearly the full temperature range of the storage. The arbitrage opportunity is thus fully identified and exploited.

Two aspects of this behavior deserve emphasis. First, the controller is causal and receives no forecast, so it cannot anticipate a price drop before observing it. This is visible in the figure, where the source switches coincide with the regime shifts rather than preceding them. The far-sightedness lies instead in the learned reaction itself. Charging the storage to full capacity during a cheap window is only justified by the expensive hours that follow, so the policy encodes a long-horizon strategy even though its execution is purely reactive. Second, the categorical and continuous outputs act in coordination. At the start of a cheap window, the gas boiler temperature setpoint drops to its lower bound, and simultaneously the EB mass flow rises to sustain both the load and the charging of the storage. The two action types jointly implement a single coherent strategy. The discrete decisions themselves are also notably clean. Each source switches only at the regime boundaries, roughly twice per day, with no chattering in between. 

\begin{figure}[h]
  \centering
  \includegraphics[width=\columnwidth]{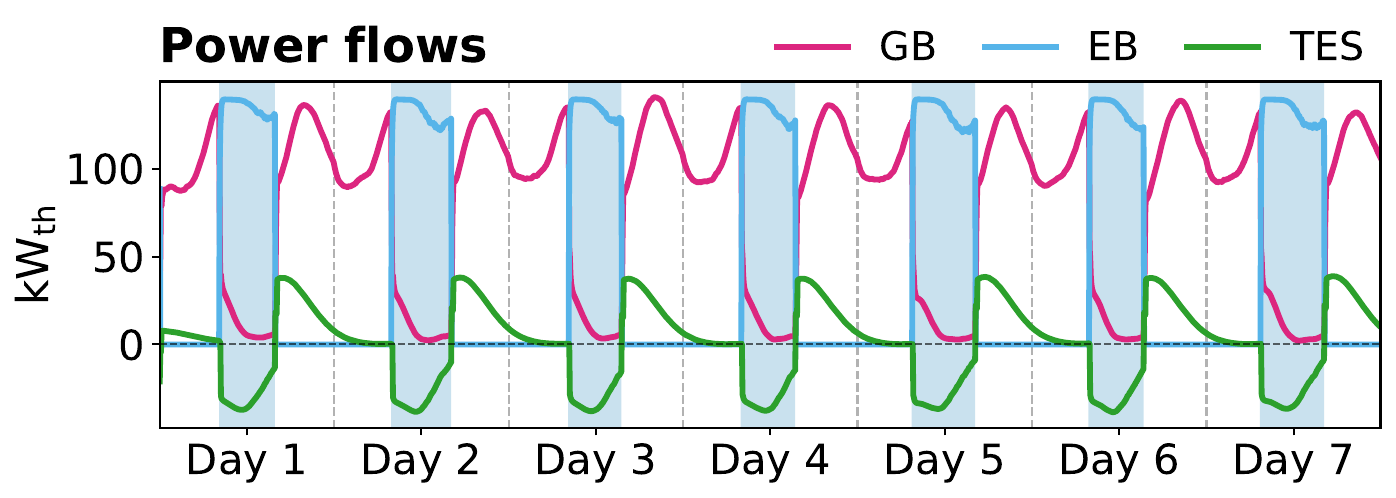}
  \caption{Power flows (Gas boiler, Electric Boiler, TES) produced by the Gumbel
  controller in closed loop on the simulator over a seven-day horizon. Shaded windows mark low-electricity-price periods, as in Fig. \ref{fig:trueprice_rollout}.}
  \label{fig:trueprice_5day}
\end{figure}

Although the three cheap windows differ in depth, width, and onset time, the response adapts to each of them rather than repeating a memorized daily schedule. This adaptability extends beyond the training horizon. Training rollouts covered two days, yet the behavior on day three is the same as that of the previous days. This is a consequence of the nature of the learned object. The controller is not an explicit plan over a fixed window but a reactive policy mapping its internal state and current measurements to actions, so nothing ties its validity to the training horizon as long as the closed-loop trajectories remain within the conditions seen during training. Figure~\ref{fig:trueprice_5day} illustrates this with a seven-day rollout, in which the thermal output profile preserves the same structure well past the two-day mark. The context awareness of the HANC is therefore highly valuable in improving its performance.

\begin{figure}[h]
  \centering
  \includegraphics[width=\columnwidth]{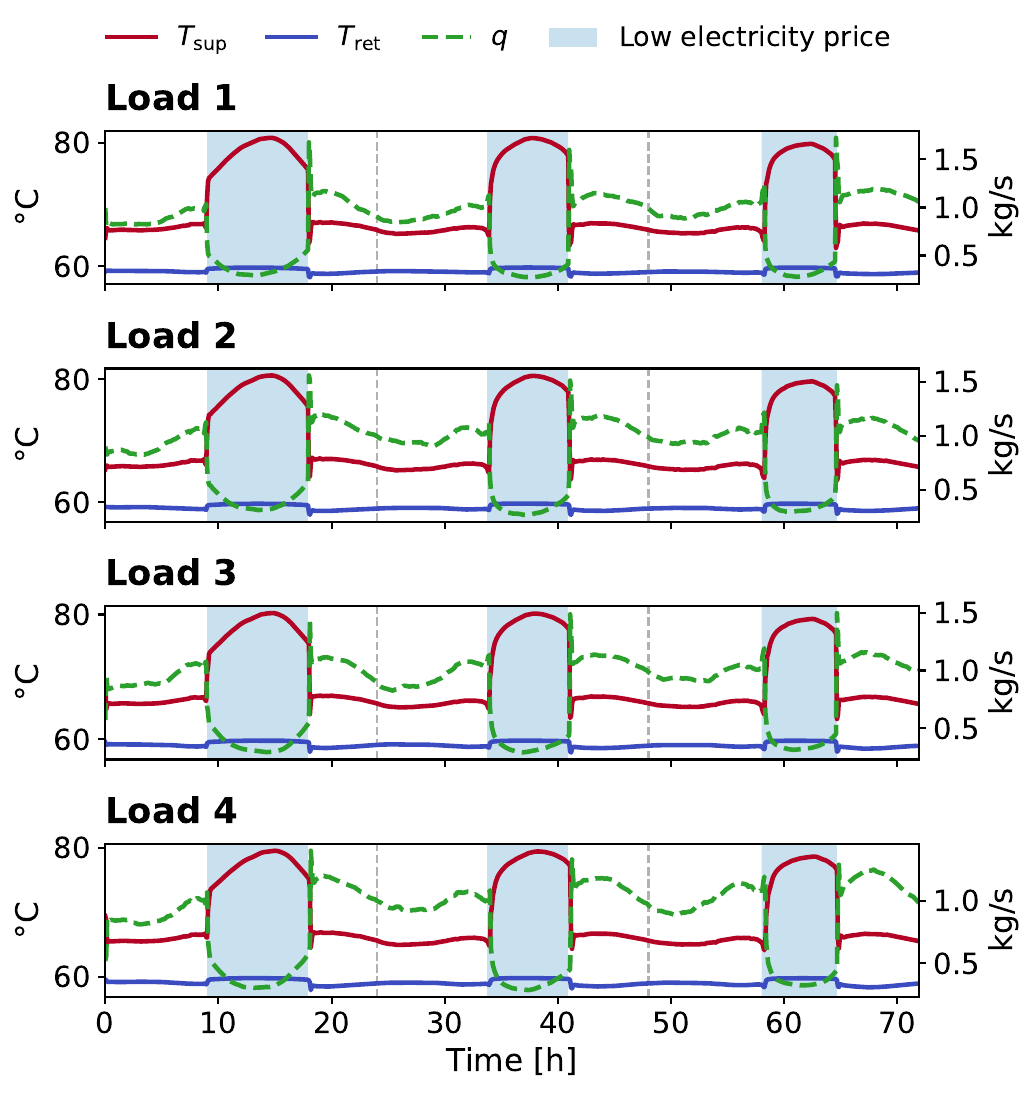}
  \caption{Supply and return temperatures and branch mass flow at each of the four heat loads.}
  \label{fig:temp_loads}
\end{figure}

Figure \ref{fig:temp_loads} shows the evolution of the DHN at the four network loads. As expected, the return temperature remains stable around $60^\circ$C due to the substations' internal controllers. The supply network alternates between two operating modes: a high-temperature, low-mass-flow mode during periods of low electricity prices, and a low-temperature, high-mass-flow mode during the remaining periods. Despite these variations, the supply temperature remains above $65^\circ$C throughout the horizon, indicating that the training successfully accounted for the operational constraints.

\subsubsection{Quantitative benchmark against rule-based strategies}

The qualitative behavior of the HANC appears promising, so it is important to assess whether these improvements translate into economic benefits. As a benchmark, we implement a representative industrial rule-based controller following the operating strategy reported in~\cite{nigro2025,LABELLA2023105429}. The benchmark is primarily clock-scheduled rather than price-scheduled. During daytime operation, the gas boiler tracks a constant supply-temperature setpoint while the electric boiler operates at constant power. During the night period (19:00--08:00), the GB supply-temperature reference is reduced and the EB is shut down. Since neither reference specifies an autonomous storage strategy, we strengthen the benchmark by making only the TES price-aware. The TES charges whenever the marginal cost of EB heat is below the GB cost and discharges otherwise, thereby exploiting energy-price arbitrage whenever it is economically beneficial. All remaining control decisions follow the fixed day/night schedule.
Table~\ref{tab:rulebased-cost} reports the closed-loop true primary-energy cost of both policies on the independent simulator under price profile imitating the real profiles with some noise, over $N=400$ rollouts with randomized load profiles and initial TES temperature. The proposed controller reduces the mean operating cost by $30\,\%$ relative to the rule-based baseline. The advantage is moreover systematic rather than average-case. The proposed controller is strictly cheaper on every one of the $400$ rollouts, with consistent per-rollout savings ranging from $28.4\%$ to $31.7\%$.

\begin{table}[h]
  \centering
  \caption{Closed-loop true primary-energy cost on the independent simulator, over $N=400$ paired
  rollouts with randomized loads and initial TES state.}
  \label{tab:rulebased-cost}
  \begin{tabular}{lcccc}
    \toprule
    Policy & Mean cost [€] & Std & vs.\ baseline  \\
    \midrule
    Proposed (Gumbel-ST)      & $280.4$ & $8.6$ & $-30.0\,\%$           \\
    Industry standard (rule-based)  & $400.4$ & $9.1$ & --   \\
    \bottomrule
  \end{tabular}
\end{table}
 
The magnitude of the gap is explained by what each policy can and
cannot react to. The baseline's only price-aware degree of freedom is
the storage mass flow. Its generation schedule is fixed by the clock,
so it keeps producing expensive EB heat through high-price hours
and cannot concentrate generation in the cheap windows, whereas the
proposed controller shifts both the source mix and the production
timing along with the storage cycle. Under the volatile price profiles,
where cheap and expensive periods are neither periodic nor aligned with
the day/night pattern, a clock schedule is systematically out of phase
with the price signal, and the cost of this misalignment compounds over
the rollout. 

\subsubsection{Gumbel versus deterministic straight-through}
\label{sec:results-gumbel-vs-ste}
 
To assess the benefit of the Gumbel noise injection, we compare the
proposed straight-through Gumbel estimator with a deterministic
straight-through variant. In the deterministic variant the controller
retains the hard forward pass and
uses the same sigmoid and softmax relaxations in the backward pass, but
the logits are passed directly to these functions without the addition
of Gumbel noise. The stochastic perturbation applied before the relaxed
discrete operations is thus the only difference between the two
estimators. To ensure a fair comparison, both controllers are trained
with identical settings, namely the same optimizer, curriculum,
annealing schedule and loss weights, and evaluated on
identical scenarios. We
deploy both methods on the simulator on $200$ samples, on price profiles which exhibit a mid-day dip like the true price profile and over a two days horizon.  

Table~\ref{tab:ste-gumbel-cost} reports the resulting operating costs.
The two methods are statistically indistinguishable. The mean costs are very close, so the deterministic estimator therefore learns an equally effective operating strategy. 
 
\begin{table}[h]
  \centering
  \caption{Closed-loop operating cost on the simulator over 200 paired
  two-day high--low--high scenarios.}
  \label{tab:ste-gumbel-cost}
  \begin{tabular}{lcc}
    \toprule
    & Deterministic ST & Gumbel ST \\
    \midrule
    Mean cost [€] & 148.7 & 146.2 \\
    Std [€] & 26.2 & 26.0 \\

    \bottomrule
  \end{tabular}
\end{table}
 
\begin{figure}[h]
  \centering
  \includegraphics[width=\columnwidth]{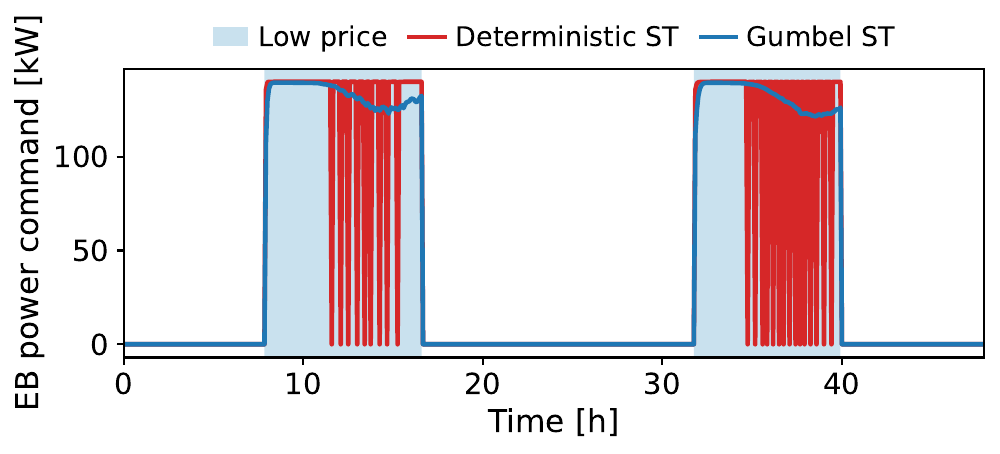}
  \caption{Closed-loop simulator rollouts of the deterministic
  straight-through (red) and Gumbel straight-through (blue) policies on
  the same two-day high--low--high scenario. Both follow the same
  economic strategy; the deterministic variant additionally toggles the
  EB gate sporadically inside the cheap-price windows.
  }
  \label{fig:ste-gumbel-traj}
\end{figure}

Where the two estimators do differ is in the treatment of the
categorical actuation itself, and the difference becomes visible in the
switching behavior of the deployed policies.
Figure~\ref{fig:ste-gumbel-traj} shows the EB power  for both controllers on the same
evaluation scenario. The schedules are nearly identical at the
timescale of the price pattern, but the deterministic variant toggles
the EB gate repeatedly within the cheap-price windows, while the
Gumbel policy performs only the four transitions the price structure
calls for. The deterministic ST controller exhibits a PWM-like switching behavior, with rapid alternation between the off and maximum-power states to approximate intermediate power levels. In contrast, the Gumbel ST controller more directly modulates the electric boiler power within its admissible operating range, resulting in more precise and sustained operating points.

Table~\ref{tab:ste-gumbel-switching} quantifies these observations across all evaluated scenarios. The largest difference appears in the EB power gate. The deterministic controller performs a median number of $51.5$ switches per two-day rollout, nearly an order of magnitude more than the Gumbel policy, whose median switch count is four. Since each price profile contains two cheap-electricity windows, the optimal strategy requires only four transitions per rollout. The HANC follows this minimal pattern on almost every sample, whereas the deterministic controller never does. Even its best-case rollout contains $24$ switches, six times more than the minimum required.

For the electric boiler mass-flow gate, both controllers never switch, instead learning to keep the valve permanently open and use it as a bypass. The two methods exhibit similar switching behavior for the TES controller, with no clear advantage for either approach.

Overall, it seems that the deterministic method induces significant and undesired switching, mainly in the EB power control. Since the two policies achieve the same cost, this chatter buys nothing and amounts to pure actuation wear. 

Further experiments were conducted with the deterministic controller, increasing the switching regularization penalty by two orders of magnitude. However, the controller still exhibited excessive jitter. This suggests that such behavior is pathological for this controller within the proposed setup.
 
\begin{table}[t]
  \centering
  \caption{Hard switching statistics per two-day simulator rollout over the
  200 paired scenarios of Table~\ref{tab:ste-gumbel-cost}.}
  \label{tab:ste-gumbel-switching}
  \begin{tabular}{llccc}
    \toprule
    Channel & Method & Min & Median & Max \\
    \midrule
    EB Power & Deterministic ST & 24 & 51.5 & 102 \\
              & Gumbel ST        & 4  & 4  & 13  \\
    \midrule
    EB Mass flow   & Deterministic ST & 0  & 0  & 0 \\
              & Gumbel ST        & 0  & 0  & 0  \\
    \midrule
    TES mode  & Deterministic ST & 4  & 4  & 14  \\
              & Gumbel ST        & 3  & 5  & 5  \\
    \bottomrule
  \end{tabular}
\end{table}

To better understand the origin of this difference, we examine the neural-network output logits at deployment, before the discrete gating operation. To the best of our knowledge, such a decision-margin analysis has not previously been reported in the context of differentiable neural control with discrete actions.

\begin{figure}[t]
  \centering
  \begin{subfigure}[t]{0.49\columnwidth}
    \centering
    \includegraphics[width=\linewidth]{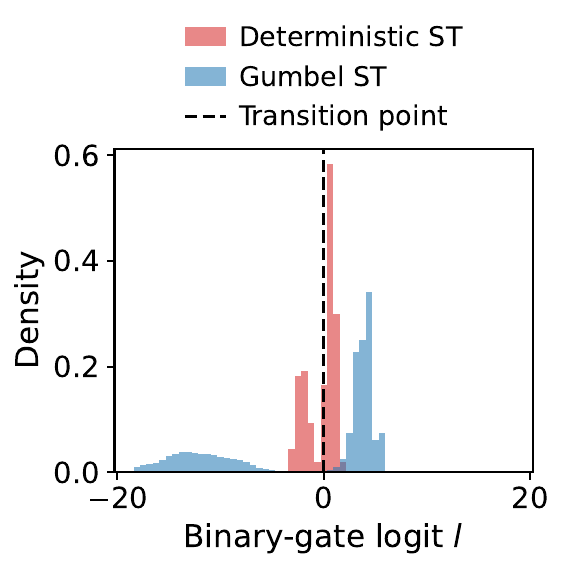}
    \caption{Binary EB gates.}
    \label{fig:logits-binary}
  \end{subfigure}\hfill
  \begin{subfigure}[t]{0.49\columnwidth}
    \centering
    \includegraphics[width=\linewidth]{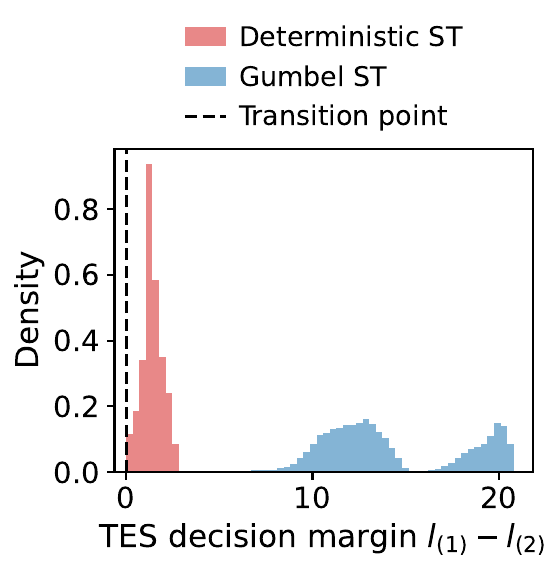}
    \caption{TES categorical head.}
    \label{fig:logits-tes}
  \end{subfigure}
  \caption{Decision-margin distributions of the two trained
  controllers. (a) Pooled signed
  logits of the two binary EB gates; the executed action flips
  at the dashed transition point $l=0$. (b) Top-two logit margin
  $l_{(1)}-l_{(2)}$ of the TES categorical head; the executed mode
  changes where the margin vanishes. Deterministic straight-through
  training leaves the logits concentrated near the transition point,
  while Gumbel training pushes them far from it.}
  \label{fig:ste-gumbel-logits}
\end{figure}

Since the executed action changes only when the decision boundary is crossed, the relevant quantity is the \emph{decision margin}: for a binary gate this is the absolute logit magnitude $|l|$ (distance to $0$), while for the TES categorical head it is the difference between the two largest logits, $l_{(1)}-l_{(2)}$. Larger margins imply that larger perturbations are required to alter the executed action. Figure~\ref{fig:ste-gumbel-logits} therefore compares the distributions of these decision margins for the two trained controllers.

The deterministic controller consistently operates close to the decision boundaries. For the binary EB gates, $48\,\%$ of the mass lies within $|l|<1$ of the switching threshold, and the median decision margin is only $1.085$. By contrast, the Gumbel-trained controller produces a strongly bimodal logit distribution whose two modes sit far from the boundary, with median margins of $3.95$ (gate on) and $12.2$ (gate off), indicating substantially more confident decisions. A similar trend is observed for the TES head. The deterministic policy produces almost exclusively margins below $3$, whereas the HANC produces a bimodal margin distribution concentrated almost entirely above $10$, with median margins of $12$ and $19.4$.

Under Gumbel training, each forward pass perturbs the logits before the discrete operation. Policies operating close to a decision boundary therefore experience stochastic changes in their categorical actions, which propagate through the rollout and increase the trajectory cost and switching penalties. The resulting optimization pressure favors separating competing logits until the perturbations no longer affect the selected action. In contrast, the deterministic estimator provides no such incentive: its hard forward pass is identical for a logit of 0.1 and 10, since both produce the same action. The optimization is therefore largely insensitive to the distance from the switching boundary, allowing solutions with smaller margins. When learning controllers with discrete decision variables, Gumbel-based training therefore offers a more robust alternative by implicitly encouraging larger decision margins and reducing the sensitivity of the learned actions to perturbations.

\section{Conclusion}

We have designed a novel learning-based control architecture that generates hybrid control actions in a differentiable way. We leveraged the Gumbel straight-through trick to achieve this, which allowed us to train these controllers using standard open-source backpropagation-through-time approaches while still guaranteeing discrete actions. Furthermore, we showed that combining the categorical and continuous control actions generated by the neural controller lets us enforce complex input constraints \textit{by design}, without compromising unconstrained optimization.

We then demonstrated the value of our method by deploying it on a complex district heating network. Although the controller was trained exclusively on a physics-informed surrogate model, it transferred successfully to a more accurate simulator representation of the network, where it identified and exploited arbitrage opportunities by producing and storing heat when electricity prices were low and discharging the TES when prices were high. The proposed controller significantly outperformed industry-standard baselines. Finally, we showed that the noise injection used during training makes the learned controller more robust to changes in operating conditions at deployment.

Given the strong performance of HANC in simulation, an interesting direction for future work is to evaluate its performance when deployed on the real plant against which the simulator was calibrated. Extending the proposed method to other applications involving categorical and hybrid control actions could also be of interest.

\bibliographystyle{IEEEtran}
\bibliography{bibliography}

@inproceedings{
jang2017categorical,
title={Categorical Reparameterization with Gumbel-Softmax},
author={Eric Jang and Shixiang Gu and Ben Poole},
booktitle={International Conference on Learning Representations},
year={2017},
url={https://openreview.net/forum?id=rkE3y85ee}
}

@inproceedings{kingma2014autoencoding,
  title     = {Auto-Encoding Variational Bayes},
  author    = {Kingma, Diederik P. and Welling, Max},
  booktitle = {International Conference on Learning Representations (ICLR)},
  year      = {2014},
  url       = {https://arxiv.org/abs/1312.6114}
}

@inproceedings{maddison2017concrete,
  title={The Concrete Distribution: A Continuous Relaxation of Discrete Random Variables},
  author={Maddison, Chris J. and Mnih, Andriy and Teh, Yee Whye},
  booktitle={International Conference on Learning Representations (ICLR)},
  year={2017},
  eprint={1611.00712},
  archivePrefix={arXiv},
  primaryClass={cs.LG}
}

@article{lottis2025minlp,
  title   = {Performance assessment of nonconvex {MINLP} formulations for district heating model predictive control},
  author  = {Lottis, Dennis and Cadenbach, Anna and H{\"a}rtel, Philipp},
  journal = {Energy},
  volume  = {335},
  pages   = {137953},
  year    = {2025},
  doi     = {10.1016/j.energy.2025.137953},
  issn    = {0360-5442}
}

@article{jansen2024minlp,
  title   = {Mixed-integer non-linear model predictive control of district heating networks},
  author  = {Jansen, Jelger and Jorissen, Filip and Helsen, Lieve},
  journal = {Applied Energy},
  volume  = {361},
  pages   = {122874},
  year    = {2024},
  doi     = {10.1016/j.apenergy.2024.122874}
}

@misc{gme_pun,
  author       = {{Gestore dei Mercati Energetici (GME)}},
  title        = {{PUN Index, Day-Ahead Market (MGP) Results}},
  howpublished = {\url{https://www.mercatoelettrico.org/en-us/Home/Results/Electricity/MGP/Results/PUN}},
  note         = {Accessed: 2026-07-10},
  year         = {2026}
}

@ARTICLE{nigro2025,
  author={Nigro, Lorenzo and La Bella, Alessio and Casella, Francesco and Scattolini, Riccardo},
  journal={IEEE Transactions on Automation Science and Engineering}, 
  title={Control-Oriented Modeling, Simulation, and Predictive Control of District Heating Networks}, 
  year={2025},
  volume={22},
  number={},
  pages={7064-7079},
  doi={10.1109/TASE.2024.3460173}}

@article{LABELLA2023105429,
title = {Optimal management and data-based predictive control of district heating systems: The Novate Milanese experimental case-study},
journal = {Control Engineering Practice},
volume = {132},
pages = {105429},
year = {2023},
issn = {0967-0661},
doi = {https://doi.org/10.1016/j.conengprac.2022.105429},
url = {https://www.sciencedirect.com/science/article/pii/S096706612200260X},
author = {Alessio {La Bella} and Ada {Del Corno}}
}

@Article{app11010455,
AUTHOR = {Buffa, Simone and Fouladfar, Mohammad Hossein and Franchini, Giuseppe and Lozano Gabarre, Ismael and Andrés Chicote, Manuel},
TITLE = {Advanced Control and Fault Detection Strategies for District Heating and Cooling Systems—A Review},
JOURNAL = {Applied Sciences},
VOLUME = {11},
YEAR = {2021},
NUMBER = {1},
ARTICLE-NUMBER = {455},
URL = {https://www.mdpi.com/2076-3417/11/1/455},
ISSN = {2076-3417},
DOI = {10.3390/app11010455}
}

@article{SLEPTSOV2021107598,
title = {Control schemes for district heating substations considering user-defined building’s indoor temperature},
journal = {Building and Environment},
volume = {191},
pages = {107598},
year = {2021},
issn = {0360-1323},
doi = {https://doi.org/10.1016/j.buildenv.2021.107598},
url = {https://www.sciencedirect.com/science/article/pii/S0360132321000135},
author = {Arseniy Sleptsov and Emanuele Crisostomi and Aldo Bischi}
}

@article{grossmann2002review,
  title={Review of nonlinear mixed-integer and disjunctive programming techniques},
  author={Grossmann, Ignacio E},
  journal={Optimization and engineering},
  volume={3},
  number={3},
  pages={227--252},
  year={2002},
  publisher={Springer}
}

@ARTICLE{6899700,
  author={Deng, Kun and Sun, Yu and Li, Sisi and Lu, Yan and Brouwer, Jack and Mehta, Prashant G. and Zhou, MengChu and Chakraborty, Amit},
  journal={IEEE Transactions on Automation Science and Engineering}, 
  title={Model Predictive Control of Central Chiller Plant With Thermal Energy Storage Via Dynamic Programming and Mixed-Integer Linear Programming}, 
  year={2015},
  volume={12},
  number={2},
  pages={565-579},
  doi={10.1109/TASE.2014.2352280}}

@Article{a15040121,
AUTHOR = {Zeile, Clemens and Weber, Tobias and Sager, Sebastian},
TITLE = {Combinatorial Integral Approximation Decompositions for Mixed-Integer Optimal Control},
JOURNAL = {Algorithms},
VOLUME = {15},
YEAR = {2022},
NUMBER = {4},
ARTICLE-NUMBER = {121},
URL = {https://www.mdpi.com/1999-4893/15/4/121},
ISSN = {1999-4893},
DOI = {10.3390/a15040121}
}

@book{rawlings2022model,
  author    = {Rawlings, James B. and Mayne, David Q. and Diehl, Moritz M.},
  title     = {Model Predictive Control: Theory, Computation, and Design},
  edition   = {2nd},
  publisher = {Nob Hill Publishing},
  year      = {2022}
}

@inproceedings{hausknecht2016deep,
  author    = {Hausknecht, Matthew and Stone, Peter},
  title     = {Deep Reinforcement Learning in Parameterized Action Space},
  booktitle = {International Conference on Learning Representations (ICLR)},
  year      = {2016},
  location  = {San Juan, Puerto Rico}
}

@inproceedings{fan2019, 
author = {Fan, Zhou and Su, Rui and Zhang, Weinan and Yu, Yong}, title = {Hybrid actor-critic reinforcement learning in parameterized action space}, year = {2019}, isbn = {9780999241141}, publisher = {AAAI Press}, booktitle = {Proceedings of the 28th International Joint Conference on Artificial Intelligence}, pages = {2279–2285}, numpages = {7}, location = {Macao, China}, series = {IJCAI'19} }

@InProceedings{pmlr-v100-neunert20a,
  title = 	 {Continuous-Discrete Reinforcement Learning for Hybrid Control in Robotics},
  author =       {Neunert, Michael and Abdolmaleki, Abbas and Wulfmeier, Markus and Lampe, Thomas and Springenberg, Tobias and Hafner, Roland and Romano, Francesco and Buchli, Jonas and Heess, Nicolas and Riedmiller, Martin},
  booktitle = 	 {Proceedings of the Conference on Robot Learning},
  pages = 	 {735--751},
  year = 	 {2020},
  editor = 	 {Kaelbling, Leslie Pack and Kragic, Danica and Sugiura, Komei},
  volume = 	 {100},
  series = 	 {Proceedings of Machine Learning Research},
  month = 	 {30 Oct--01 Nov},
  publisher =    {PMLR},
  url = 	 {https://proceedings.mlr.press/v100/neunert20a.html}
}

@article{xiong2018parametrized,
  title={Parametrized deep q-networks learning: Reinforcement learning with discrete-continuous hybrid action space},
  author={Xiong, Jiechao and Wang, Qing and Yang, Zhuoran and Sun, Peng and Han, Lei and Zheng, Yang and Fu, Haobo and Zhang, Tong and Liu, Ji and Liu, Han},
  journal={arXiv preprint arXiv:1810.06394},
  year={2018}
}

@article{Bester2019MultiPassQF,
  title={Multi-Pass Q-Networks for Deep Reinforcement Learning with Parameterised Action Spaces},
  author={Craig J. Bester and Steven James and George Dimitri Konidaris},
  journal={ArXiv},
  year={2019},
  volume={abs/1905.04388},
  url={https://api.semanticscholar.org/CorpusID:152282275}
}

@article{stepanovic2022rl,
  author  = {Stepanovic, Ksenija and Wu, Jichen and Everhardt, Rob and de Weerdt, Mathijs},
  journal = {Energies},
  title   = {Unlocking the Flexibility of District Heating Pipeline Energy Storage with Reinforcement Learning},
  year    = {2022},
  volume  = {15},
  number  = {9},
  pages   = {3290},
  doi     = {10.3390/en15093290}
}

@article{han2023rainbow,
  author  = {Han, Gwangwoo and Joo, Hong-Jin and Lim, Hee-Won and An, Young-Sub and Lee, Wang-Je and Lee, Kyoung-Ho},
  journal = {Energy},
  title   = {Data-driven heat pump operation strategy using rainbow deep reinforcement learning for significant reduction of electricity cost},
  year    = {2023},
  volume  = {270},
  pages   = {126913},
  doi     = {10.1016/j.energy.2023.126913}
}

@article{GONG2024107672,
title = {Intelligent control of district heating system based on RDPG},
journal = {Engineering Applications of Artificial Intelligence},
volume = {129},
pages = {107672},
year = {2024},
issn = {0952-1976},
doi = {https://doi.org/10.1016/j.engappai.2023.107672},
url = {https://www.sciencedirect.com/science/article/pii/S0952197623018560},
author = {Mingju Gong and Yan Liu and Jiawang Sun and Wei Xu and Wenxiang Li and Changcheng Yan and Wencheng Fu}
}

@article{DRGONA202280,
title = {Differentiable predictive control: Deep learning alternative to explicit model predictive control for unknown nonlinear systems},
journal = {Journal of Process Control},
volume = {116},
pages = {80-92},
year = {2022},
issn = {0959-1524},
doi = {https://doi.org/10.1016/j.jprocont.2022.06.001},
url = {https://www.sciencedirect.com/science/article/pii/S0959152422000981},
author = {Ján Drgoňa and Karol Kiš and Aaron Tuor and Draguna Vrabie and Martin Klaučo}
}

@inproceedings{xu2022shac,
  author    = {Xu, Jie and Makoviychuk, Viktor and Narang, Yashraj and Ramos, Fabio and Matusik, Wojciech and Garg, Animesh and Macklin, Miles},
  booktitle = {International Conference on Learning Representations (ICLR)},
  title     = {Accelerated Policy Learning with Parallel Differentiable Simulation},
  year      = {2022}
}

@ARTICLE{pb,
  author={Furieri, Luca and Galimberti, Clara Lucía and Ferrari-Trecate, Giancarlo},
  journal={IEEE Open Journal of Control Systems}, 
  title={Learning to Boost the Performance of Stable Nonlinear Systems}, 
  year={2024},
  volume={3},
  number={},
  pages={342-357},
  doi={10.1109/OJCSYS.2024.3441768}
}

@INPROCEEDINGS{rpb,
  author={Kirsch, Nicolas and Massai, Leonardo and Ferrari-Trecate, Giancarlo},
  booktitle={2025 IEEE 64th Conference on Decision and Control (CDC)}, 
  title={Boosting the transient performance of reference tracking controllers with neural networks}, 
  year={2025},
  volume={},
  number={},
  pages={8015-8021},
  doi={10.1109/CDC57313.2025.11312988}}

@article{boldocky2026data,
  title={Data center chiller plant optimization via mixed-integer nonlinear differentiable predictive control},
  author={Boldock{\`y}, J{\'a}n and Faulkner, Cary and Michael, Elad and Gulan, Martin and Tuor, Aaron and Drgo{\v{n}}a, J{\'a}n},
  journal={Control Engineering Practice},
  volume={174},
  pages={107063},
  year={2026},
  publisher={Elsevier}
}

@article{boldocky2025learning,
  title={Learning to solve parametric mixed-integer optimal control problems via differentiable predictive control},
  author={Boldock{\`y}, J{\'a}n and Javan, Shahriar Dadras and Gulan, Martin and M{\"o}nnigmann, Martin and Drgo{\v{n}}a, J{\'a}n},
  journal={arXiv preprint arXiv:2506.19646},
  year={2025}
}

@article{bengio2013estimating,
  title={Estimating or propagating gradients through stochastic neurons for conditional computation},
  author={Bengio, Yoshua and L{\'e}onard, Nicholas and Courville, Aaron},
  journal={arXiv preprint arXiv:1308.3432},
  year={2013}
}

@article{MANCARELLA20141,
title = {MES (multi-energy systems): An overview of concepts and evaluation models},
journal = {Energy},
volume = {65},
pages = {1-17},
year = {2014},
issn = {0360-5442},
doi = {https://doi.org/10.1016/j.energy.2013.10.041},
url = {https://www.sciencedirect.com/science/article/pii/S0360544213008931},
author = {Pierluigi Mancarella}
}

@article{blanco2019novel,
  author  = {Blanco, Ignacio and Andersen, Anders N. and Guericke, Daniela and Madsen, Henrik},
  title   = {A novel bidding method for combined heat and power units in district heating systems},
  journal = {Energy Systems},
  volume  = {11},
  pages   = {1137--1156},
  year    = {2020},
  doi     = {10.1007/s12667-019-00352-0}
}

@article{hering2022operational,
  author  = {Hering, Dominik and Faller, Michael R. and Xhonneux, André and Müller, Dirk},
  title   = {Operational optimization of a 4th generation district heating network with mixed integer quadratically constrained programming},
  journal = {Energy},
  volume  = {250},
  pages   = {123766},
  year    = {2022},
  doi     = {10.1016/j.energy.2022.123766}
}

@inproceedings{alvarado2024development,
  title={Development and Experimental Validation of an Open-Source Model Library for District Heating Network Simulation},
  author={Alvarado, Marcelo Andre Muro and Anderis, Claudio and Lazzari, Riccardo and Nigro, Lorenzo and La Bella, Alessio},
  booktitle={2024 Open Source Modelling and Simulation of Energy Systems (OSMSES)},
  pages={1--6},
  year={2024},
  organization={IEEE}
}

@article{nigro2025experimental,
  title={Experimental validation and design of nonlinear predictive control and moving horizon estimation for district heating},
  author={Nigro, Lorenzo and Polimeni, Simone and Anderis, Claudio and Muro, Marcelo and La Bella, Alessio},
  journal={IFAC-PapersOnLine},
  volume={59},
  number={9},
  pages={157--162},
  year={2025},
  publisher={Elsevier}
}

@inproceedings{sgadari_2026_topology_pil,
  author    = {Sgadari, Corrado and Bianchi, Federico and Polimeni, Simone and La Bella, Alessio},
  title     = {Topology-guided physics-informed learning of district heating networks with guaranteed conservation laws},
  booktitle = {Proceedings of the 23rd IFAC World Congress},
  year      = {2026},
  address   = {Busan, Republic of Korea},
  note      = {In press}
}

@article{deGiuli2024,
  author  = {{Boca de Giuli}, Laura and La Bella, Alessio and Scattolini, Riccardo},
  title   = {Physics-Informed Neural Network Modeling and Predictive Control of District Heating Systems},
  journal = {IEEE Transactions on Control Systems Technology},
  year    = {2024},
  volume  = {32},
  number  = {4},
  pages   = {1182--1195},
  doi     = {10.1109/TCST.2024.3355476}
}

@article{sibeijn2025,
  author  = {Sibeijn, Max and Ahmed, Saeed and Khosravi, Mohammad and Keviczky, Tam{\'a}s},
  title   = {Economic Nonlinear Model Predictive Control of Prosumer District Heating Networks},
  journal = {IEEE Transactions on Control Systems Technology},
  year    = {2025},
  volume  = {33},
  number  = {5},
  pages   = {1879--1894},
  doi     = {10.1109/TCST.2025.3561501}
}

@article{krug2021,
  author  = {Krug, Richard and Mehrmann, Volker and Schmidt, Martin},
  title   = {Nonlinear optimization of district heating networks},
  journal = {Optimization and Engineering},
  year    = {2021},
  volume  = {22},
  number  = {2},
  pages   = {783--819},
  doi     = {10.1007/s11081-020-09549-0}
}

@article{taylor2021,
  author  = {Taylor, Michael and Long, Sebastian and Marjanovic, Ognjen and Parisio, Alessandra},
  title   = {Model Predictive Control of Smart Districts With Fifth Generation Heating and Cooling Networks},
  journal = {IEEE Transactions on Energy Conversion},
  year    = {2021},
  volume  = {36},
  number  = {4},
  pages   = {2659--2669},
  doi     = {10.1109/TEC.2021.3082405}
}

@article{taieb2025,
  author  = {Taieb, Zinedine and Sandou, Guillaume and Vasset, Nicolas and Gaoua, Yacine and Pichenot, Gr{\'e}goire},
  title   = {Approximate explicit model predictive control approach for the unit commitment production of a district heating network},
  journal = {IFAC-PapersOnLine},
  year    = {2025},
  volume  = {59},
  number  = {12},
  pages   = {25--30},
  doi     = {10.1016/j.ifacol.2025.09.562}
}

@inproceedings{okamoto2026accelerating,
  author    = {Okamoto, Morimasa and Wasa, Yasuaki},
  title     = {Accelerating {MINLP}-Based District Cooling Operational
               Planning Using Neural Network Controller},
  booktitle = {Preprints of the 23rd IFAC World Congress},
  address   = {Busan, Republic of Korea},
  month     = aug,
  year      = {2026},
  pages     = {3050--3055},
  note      = {Accepted for publication in IFAC-PapersOnLine},
}

@misc{zheng2026uphes,
  author        = {Zheng, Honghui and Boldock{\'y}, J{\'a}n and Dvorkin, Yury and Drgo{\v{n}}a, J{\'a}n},
  title         = {Mixed-Integer Nonlinear Differentiable Predictive Control for Underground Pumped Hydro Energy Storage Systems},
  year          = {2026},
  eprint        = {2609.17964},
  archivePrefix = {arXiv},
  primaryClass  = {eess.SY},
  note          = {Accepted for publication at the 65th IEEE Conference on Decision and Control (CDC)}
}

@ARTICLE{58337,
  author={Werbos, P.J.},
  journal={Proceedings of the IEEE}, 
  title={Backpropagation through time: what it does and how to do it}, 
  year={1990},
  volume={78},
  number={10},
  pages={1550-1560},
  doi={10.1109/5.58337}}

\end{document}